\documentstyle[preprint,eqsecnum,aps,epsf]{revtex}
\begin{document}
\tightenlines   
\draft
\title{%
Nambu--Jona-Lasinio Model Coupled to Constant\\
Electromagnetic Fields in $D$-Dimension\thanks{KYUSHU-HET-40}
}
%
\author{%
Masaru~ISHI-I\thanks{isii1scp@mbox.nc.kyushu-u.ac.jp},
Taro~KASHIWA\thanks{taro1scp@mbox.nc.kyushu-u.ac.jp}, and
Naoki~TANIMURA\thanks{tnmr1scp@mbox.nc.kyushu-u.ac.jp}
}
\address{%
Department of Physics, Kyushu University, 
Fukuoka 812-81, JAPAN 
}
\date{\today}
\maketitle
%
%
%
\begin{abstract}
Critical dynamics of the Nambu--Jona-Lasinio model, 
coupled to a constant electromagnetic field in $D=2$, $3$, and $4$, 
is reconsidered from a
viewpoint of infrared behavior and vacuum instability. 
The latter is associated with constant 
electric fields and cannot be avoidable in the nonperturbative 
framework obtained through the proper time method. 
As for magnetic fields, 
an infrared cut-off is essential to investigate the critical 
phenomena. The result reconfirms the fact 
that the critical coupling in $D=3$ and $4$ goes to zero even under an 
infinitesimal magnetic field. 
There also shows that a non-vanishing $F_{\mu\nu}\widetilde F^{\mu\nu}$ 
causes instability. A perturbation with respect to external fields 
is adopted to investigate critical quantities, but the resultant 
asymptotic expansion excellently matches with the exact value. 
\end{abstract}
%
%
\section{Introduction}\label{intro}
The four-fermi interaction model by Nambu and Jona-Lasinio 
(NJL) \cite{NJL} has been discussed to 
investigate the dynamical symmetry breaking (DSB) in a number of cases 
in two, three, and four dimensions. Especially interesting situations are 
found such that NJL is coupled to external sources, which enables us 
to peep into detailed structures of DSB and obtain information on 
the chiral symmetry breaking ($\chi$SB) in the QCD vacuum, the planar 
($2+1$-dimensional) dynamics in solid state physics, or the early 
universe when coupled to a curved space-time \cite{IMO}. In this 
respect  
the NJL model minimally coupled to the electromagnetic fields is 
discussed by many authors to explain how $\chi$SB 
could be changed under the influence of the 
electromagnetic fields: Klevansky and Lemmer \cite{KL} find that 
a pure electric field opposes $\chi$SB to restore chiral symmetry, 
meanwhile a pure magnetic field enhances $\chi$SB. 
The former result has been generalized to non-abelian gauge fields by 
Suganuma and Tatsumi, where they argue about chiral symmetry 
restoration by (color) electric fields\cite{ST}. Meanwhile the 
latter result is further investigated by Gusynin, Miransky, and Shovkovy 
who find that there occurs the mass generation even at the weakest 
attractive interaction in $2+1$-dimension \cite{GMS1} ($\chi$SB in 
$2+1$-dimension can be realized as a flavor symmetry breaking by 
introducing an additional fermion \cite{ABKW}) and in $3+1$-dimension 
\cite{GMS2} and emphasize it by means of the dimensional 
reduction. This implies that 
{\em the critical coupling is zero even if 
the applying magnetic field is infinitesimal}, which might however 
contradict with a na\"{\i}ve consideration. 
The motivation for this work lies here. 

Our strategy to this issue is: to start with the Euclidean 
path integral 
representation of the NJL model minimally coupled to constant 
electromagnetic fields. Following the standard procedure, that is, 
introducing auxiliary fields and integrating with respect to the 
fermion field, we arrive at the pure  bosonic path integral consisting 
of the auxiliary fields and a functional determinant. We then rely on 
the semi-classical approximation and employ the Fock-Schwinger proper 
time method \cite{IZ} in order to define a functional determinant. 
Although the proper time method can automatically provide a 
gauge-invariant 
ultraviolet (UV) regularization in terms of the gamma or zeta function 
similar to the dimensional 
regularization, we introduce a cut-off $\Lambda^{2}$ to grasp physical 
situations better, which, however, still be gauge invariant as well as 
Lorentz covariant. Moreover {\em an infrared (IR)  
cut-off $\epsilon$ must also be introduced}, 
since infrared divergences arise when external 
fields are coupled to a massless state.  So far a very little care has 
been devoted to this fact but it is inevitable in order to 
discuss effects of electromagnetic fields to the NJL model; since we 
are interested in the transition from massless to massive states or 
vice versa 
under an influence of external fields. 
From these procedures we obtain the effective potential. 

Another issue comes up when an electric field comes into play. The 
effective potential becomes complex, whose imaginary part implies 
the vacuum 
instability because of the creation of fermion and antifermion 
pairs. (The phenomenon is closely related to the Klein paradox in the 
one body Dirac equation \cite{GMR}.) However if we notice that the 
imaginary part behaves as ${\mathrm e}^{-m^{2}/E}$ with $E$ being the 
magnitude of electric field, it is negligible when 
\begin{equation}
E<m^{2}\ ,
\label{eltm}
\end{equation}
which leads us to the situation that the magnitude of electric fields 
must always be infinitesimal. However it is enough to live in this 
world, since we are interested in a change of a vacuum 
given by the NJL model to that with external fields, 
so that we can consider them so small as perturbations. 
With these spirits, we find that there is no notion of criticality in 
the pure electric field case, since that is defined through a 
transition from a massive to a massless state but there is not any state 
with a mass less than a magnitude of the electric field owing to 
(\ref{eltm}). On the contrary, in a pure magnetic field case, we find 
critical couplings. They are nonvanishing as far as the infrared 
cut-off $\epsilon$ is kept finite but eventually become zero when 
$\epsilon\rightarrow0$. 
This reconfirms the results of Gusynin, Miransky, and Shovkovy 
who have relied on the argument in terms of the dimensional reduction. 
These results are obtained through perturbations with respect to 
external fields but a resultant asymptotic expansion is found well 
matched with the exact value even if the first few terms are adopted. 
This observation would be interesting. 

The paper is organized as follows: in Sec. \ref{general}, we develop a 
general formalism for computing an effective potential under constant 
electromagnetic fields. In the subsequent Sec. \ref{electric}, 
\ref{magnetic}, and \ref{electromagnetic}, 
a pure electric, magnetic, and 
a non-vanishing $F_{\mu\nu}\widetilde F^{\mu\nu}$ ($\equiv 
\epsilon^{\mu\nu\rho\sigma}F_{\mu\nu}F_{\rho\sigma}/2$) in $D=4$ cases are 
discussed in order. The final Sec. \ref{conclusion} is devoted to conclusion. 
In Appendix \ref{appendixa}, calculations of the trace and the 
determinant in deriving the effective potential are given. 

\section{General Formulation}\label{general}

In this section we describe the model and develop a 
general formulation for obtaining the one-loop effective potential of the 
$D$-dimensional NJL model. The Lagrangian for the 
NJL model minimally coupled to external electromagnetic fields is 
\begin{equation} 
{\cal L}=-\overline\psi\Bigl\{\gamma_{\mu}
\bigl(\partial_{\mu}-iA_{\mu}\bigr)\Bigr\}\psi
+{g^{2}\over2}\left\{\begin{array}{ll}
\left[\bigl(\overline\psi\psi\bigr)^{2}
+\bigl(\overline\psi i\gamma_{5}\psi\bigr)^{2}\right]\ ,&D=2,4\ ,\\ 
\noalign{\vspace{1ex}}
\left[\bigl(\overline\psi\psi\bigr)^{2}
+\bigl(\overline\psi i\gamma_{4}\psi\bigr)^{2}
+\bigl(\overline\psi i\gamma_{5}\psi\bigr)^{2}\right]\ ,&D=3\ ,
\end{array}\right.
\label{NJLL}
\end{equation}
where the electromagnetic coupling 
constant has been absorbed into the definition of $A_{\mu}$. Apart from 
a usual $4$-dimensional case, 
$\psi$ is a two-component spinor 
with gamma matrices 
\begin{equation}
\gamma_{\mu}=\sigma_{\mu}\,,\quad\gamma_{5}
=-i\gamma_{1}\gamma_{2}\,,\quad
\sigma_{\mu\nu}\equiv[\gamma_{\mu},\gamma_{\nu}]
/2i=\epsilon_{\mu\nu}\sigma_{3}\,;\quad\mu=1,2\,,
\end{equation}
in $2$-dimension. 
For the $3$-dimensional case, a spinorial representation of the 
Lorentz group is given by two-component spinors, so that 
corresponding gamma matrices are $2\times2$. There is no 
chiral symmetry. 
In order to be able to discuss chiral symmetry, we introduce another 
flavor such that \cite{ABKW} 
\begin{equation}
\psi=\left(\begin{array}{c}\psi_{1}\\\psi_{2}\end{array}\right)\ ,\quad
\overline\psi\equiv\left(\begin{array}{cc}\overline\psi_{1}
&\overline\psi_{2}\end{array}\right)
\equiv\left(\begin{array}{cc}\psi_{1}^{\dag}\sigma_{3}
&\psi_{2}^{\dag}\sigma_{3}
\end{array}\right)\ ,
\end{equation}
and $4\times4$ gamma matrices
\begin{equation}
\gamma_{\mu}=\left(\begin{array}{cc}\sigma_{\mu}&0\\
0&-\sigma_{\mu}\end{array}\right)\,;\mu=1\sim3\,,\quad
\gamma_{4}=\left(\begin{array}{cc}0&{\bbox1}\\{\bbox1}&0\end{array}\right)\,,
\quad\gamma_{5}=\gamma_{1}\gamma_{2}\gamma_{3}\gamma_{4}
=\left(\begin{array}{cc}0&i{\bbox1}\\-i{\bbox1}&0\end{array}\right)\ .
\label{gamma3d}
\end{equation}
The chiral symmetry realizes as 
\begin{equation}
\psi\longrightarrow{\rm e}^{i\alpha\gamma_{4}}\psi\,,\quad
\psi\longrightarrow{\rm e}^{i\beta\gamma_{5}}\psi\ ,
\end{equation}
yielding a global $U(2)$ symmetry which is broken by a mass 
term into $U(1)\times U(1)$. 

The partition function of the model is read as 
\begin{eqnarray}
\lefteqn{Z[A]\equiv\int\!\!{\mathcal D}\psi{\mathcal D}\overline\psi\exp
\left[\!\int\!\!d^{D}x\,{\cal L}\right]}
\nonumber \\ 
\noalign{\vskip -1ex}	
&&\\
\noalign{\vskip -1ex}  
&=&\int\!\!{\mathcal D}\sigma{\mathcal D}{\bbox\pi}{\mathcal D}\psi
{\mathcal D}\overline\psi\exp\left[-\!\int\!\!d^{D}x
\Bigl[{1\over2g^{2}}\bigl(\sigma^{2}+{\bbox\pi}^{2}\bigr)
+\overline\psi\Bigl\{\gamma_{\mu}\bigl(\partial_{\mu}-iA_{\mu}\bigr)
+\bigl(\sigma+i{\bbox{\pi\cdot{\mit\Gamma}}}\bigr)\Bigr\}\psi\Bigr]\right]\ ,\nonumber 
\end{eqnarray}
where auxiliary fields, $\sigma$ and ${\bbox\pi}$, have 
been introduced as usual; 
\begin{equation}
{\bbox{\pi\cdot{\mit\Gamma}}}=\left\{
\begin{array}{ll}\pi\gamma_{5}\,&\mbox{for }D=2,4\,,\\
\pi_{1}\gamma_{4}+\pi_{2}\gamma_{5}\,&\mbox{for }D=3\ .
\end{array}\right.
\end{equation}
The fermionic integration gives the 
functional determinant: 
\begin{eqnarray}
\lefteqn{\int\!\!{\mathcal D}\psi{\mathcal D}\overline\psi\exp
\left[-\int\!\!d^{D}x\,\overline\psi
\Bigl\{\gamma_{\mu}\bigl(\partial_{\mu}-iA_{\mu}\bigr)
+\bigl(\sigma+i{\bbox{\pi\cdot{\mit\Gamma}}}\bigr)\Bigr\}\psi\right]}
\nonumber\\
&&\hspace{15em}\equiv
{\mathrm{Det}}\Bigl[\gamma_{\mu}\bigl(\partial_{\mu}-iA_{\mu}\bigr)
+\bigl(\sigma+i{\bbox{\pi\cdot{\mit\Gamma}}}\bigr)\Bigr]\ .
\label{det}
\end{eqnarray}
We then perform the semiclassical approximation, 
that is, shift, $\sigma\rightarrow m+\sigma'$, 
${\bbox\pi}\rightarrow{\bbox\pi}'$, 
and assign 
$\sigma'$ and  ${\bbox\pi}'$ as the new integration variables to find  
\begin{equation}
Z[A]=\exp\left[-VT{\mbox{\Large$v$}}_{D}(m)\right]   
\Bigl(1+O(\mbox{$2$-loop})\Bigr)\ ,
\end{equation}
with 
\begin{equation}
\mbox{\Large$v$}_{D}(m)\equiv{m^{2}\over2g^{2}}-{1\over VT}
\ln{\mathrm{Det}}\Bigl[\gamma_{\mu}
\bigl(\partial_{\mu}-iA_{\mu}\bigr)+m\Bigr]\ ,
\label{vdm}
\end{equation}
where $V$ is the $(D-1)$-dimensional volume of the system 
and $T$ is the Euclidean time 
interval. (This semiclassical approximation would be more justified by 
introducing $N$ fermion pieces and taking $N\rightarrow\infty$.) 
It should be understood 
that the terms of $O(\mbox{2-loop})$ 
are given by integrations with respect to $\sigma'$ and ${\bbox\pi}'$. 
${\mbox{\Large$v$}}_{D}(m)$ in (\ref{vdm}) is the one-loop effective 
potential from which we 
can see the phase structure of the model. 

In order to make (\ref{det}) well defined, we normalize it as 
follows: 
\begin{eqnarray}
I_{D}&\equiv&\ln\left[{{\mathrm{Det}}
[\gamma_{\mu}(\partial_{\mu}-iA_{\mu})+m]
\over{\mathrm{Det}}[\gamma_{\mu}\partial_{\mu}]}\right]
\nonumber\\
&=&{\mathrm{Tr}}{\mathrm{Ln}}\Bigl(\gamma_{\mu}
\bigl(\partial_{\mu}-iA_{\mu}\bigr)+m\Bigr)
-{\mathrm{Tr}}{\mathrm{Ln}}\Bigl(\gamma_{\mu}\partial_{\mu}\Bigr)
\nonumber\\
&=&{1\over2}{\mathrm{Tr}}{\mathrm{Ln}}
\Bigl(-\bigl(\partial_{\mu}-iA_{\mu}\bigr)^{2}
-{1\over2}\sigma_{\mu\nu}F_{\mu\nu}+m^{2}\Bigr)
-{1\over2}{\mathrm{Tr}}{\mathrm{Ln}}\Bigl(-\partial_{\mu}^{2}\Bigr)\ ,
\label{id}
\end{eqnarray}
where the trace operation, designated by ${\mathrm{Tr}}$,
must be taken with respect to the 
space-time as well as the gamma matrices, whereas ${\mathrm{tr}}$ 
implies that only for the gamma matrices. 
With the use of the identity, 
\begin{equation}
\ln H=-\lim_{s\rightarrow0}\left[\int_{0}^{\infty}\!d\tau\,\tau^{s-1}
{\mathrm e}^{-\tau H}-\Gamma(s)\right]\ ,
\end{equation}
$I_{D}$ can be rewritten as 
\begin{eqnarray}
I_{D}&=&-{1\over2}\lim_{s\rightarrow0}\int_{0}^{\infty}\!d\tau\,\tau^{s-1}
{\mathrm{Tr}}\left({\mathrm e}^{-\tau[-(\partial_{\mu}-iA_{\mu})^{2}
-{1\over2}\sigma_{\mu\nu}F_{\mu\nu}+m^{2}]}
-{\mathrm e}^{-\tau[-\partial_{\mu}^{2}]}\right)
\nonumber\\
&=&-{1\over2}\lim_{s\rightarrow0}\int_{0}^{\infty}\!d\tau\,\tau^{s-1}
\Biggl[{\mathrm e}^{-\tau m^{2}}{\mathrm{tr}}
({\mathrm e}^{{\tau\over2}\sigma_{\mu\nu}F_{\mu\nu}})\int\!\!d^{D}x
\langle x\vert{\mathrm e}^{\tau[(\partial_{\mu}-iA_{\mu})^{2}]}
\vert x\rangle
\nonumber\\
&&-{\mathrm{tr}}{\bf 1}\int\!\!d^{D}x
\langle x\vert{\mathrm e}^{\tau\partial_{\mu}^{2}}\vert x\rangle\Biggr]\ ,
\label{ird}
\end{eqnarray}
where, in the second line, $F_{\mu\nu}$ has been assumed to be constant 
and $\sigma_{\mu\nu}\equiv\bigl[\gamma_{\mu},\gamma_{\nu}\bigr]/2i$. The 
kernel $\langle x\vert{\mathrm e}^{\tau[(\partial_{\mu}-iA_{\mu})^{2}]}
\vert x'\rangle$ with a constant $F_{\mu\nu}$ 
can be calculated by means of the proper 
time method \cite{IZ} as 
\begin{eqnarray}
\langle x\vert{\mathrm e}^{\tau[(\partial_{\mu}-iA_{\mu})^{2}]}
\vert x'\rangle
&=&{1\over(4\pi\tau)^{D/2}}\exp
\left[i\int_{x'}^{x}\!d\xi_{\mu}A_{\mu}(\xi)\right]
\left[\det\left({\sin(\tau F)\over\tau F}\right)\right]^{-{1\over2}}
\nonumber\\
&&\hspace{30pt}\times\exp\left[-{1\over4\tau}(x-x')_{\mu}
(\tau F\cot\tau F)_{\mu\nu}(x-x')_{\nu}\right]\ ,
\label{kernel}
\end{eqnarray}
where $F$ denotes a $D\times D$ matrix whose components are $F_{\mu\nu}$. 
Combining (\ref{ird}) and (\ref{kernel}) we obtain 
\begin{equation}
I_{D}=-VT\lim_{s\rightarrow0}{{\mathrm{tr}}{\bf1}\over2(4\pi)^{D/2}}
\int_{0}^{\infty}\!d\tau\,\tau^{s-{D\over2}-1}\left\{
{\mathrm e}^{-\tau m^{2}}G_{D}(\tau F)-1\right\}\ .
\label{I}
\end{equation}
where $G_{D}(\tau F)$ reads 
\begin{eqnarray}
G_{D}(\tau F)&\equiv&
{{\mathrm{tr}}({\mathrm e}^{{\tau\over2}\sigma_{\mu\nu}{F}_{\mu\nu}})
\over{\mathrm{tr}}{\bf1}}
\left[\det\left({\sin(\tau{F})\over\tau{F}}\right)
\right]^{-{1\over2}}
\nonumber\\
&=&\left\{
\begin{array}{lcl}
\tau{F}_{D}\coth(\tau{F}_{D})&{\mathrm{for}}&D=2,3\ ,\\
\noalign{\vspace{1ex}}
\tau^{2}{F}_{+}{F}_{-}\coth(\tau{F}_{+})\coth(\tau{F}_{-})
&{\mathrm{for}}&D=4\ ,
\end{array}\right.
\label{trdet}
\end{eqnarray}
with  
\begin{equation}
\begin{array}{lcl}
{F}_{2}&=&E\ ,\\
{F}_{3}&=&\sqrt{B^{2}+{\bbox E}^{2}}\ ,\\
{F}_{\pm}&=&\left\{\vert{\bbox{B+E}}\vert\pm
\vert{\bbox{B-E}}\vert\right\}/2\ .
\end{array}\label{Fs}
\end{equation}
(Details are shown in Appendix \ref{appendixa}.) 
As was mentioned above 
${\mathrm{tr}}{\bf1}$ in (\ref{I}) and (\ref{trdet}) 
is $2^{D/2}$ for $D=2,4$ and $2^{(D+1)/2}$ for $D=3$. 
Although the integral (\ref{I}) has entirely 
been regularized if an analytic continuation is made for $s$, 
in order to grasp a physical situation better, 
an UV cut-off $\Lambda^{2}$ is introduced as is done 
in the ordinary gap equation \cite{NJL}. Moreover, an IR 
cut-off $\epsilon$ is indispensable, 
since if $m^{2}=0$ and $F\ne0$, the integral (\ref{I}) becomes 
divergent. Therefore we consider instead of (\ref{I}) 
\begin{equation}
{I_{D}}^{\mathrm r}\equiv-VT{{\mathrm{tr}}{\bf1}\over2(4\pi)^{D/2}}
\int_{1/\Lambda^{2}}^{\infty}\!d\tau\,\tau^{-{D\over2}-1}
\left\{{\mathrm e}^{-\tau(m^{2}+\epsilon)}G_{D}(\tau F)-1\right\}\ .
\label{Ir}
\end{equation}
This regularized ${{I}_{D}}^{\mathrm r}$ enables us to 
investigate dynamics 
near the massless region, so does that of $\chi$SB. 
Now the well-defined effective potential is given by 
\begin{equation}
{\mbox{\Large$v$}}_{D}(m)={m^{2}\over2g^{2}}
+{{\mathrm{tr}}{\bf1}\over2(4\pi)^{D/2}}
\int_{1/\Lambda^{2}}^{\infty}\!d\tau\,\tau^{-{D\over2}-1}
\left\{{\mathrm e}^{-\tau(m^{2}+\epsilon)}G_{D}(\tau F)-1\right\}\ .
\end{equation}

We then introduce dimensionless 
quantities, $x\equiv(m^{2}+\epsilon)/\Lambda^{2}$ and 
${\cal F}\equiv F/\Lambda^{2}$ 
$({\cal F}_{\mu\nu}\equiv F_{\mu\nu}/\Lambda^{2})$, obeying 
$\epsilon/\Lambda^{2}\le x<1$ and $\vert{\cal F}_{\mu\nu}\vert<1$. 
(Recall that mass dimension of the gauge field 
is always one because of inclusion of the coupling constant.) 
Therefore the (dimensionless) effective potential 
$\overline{\mbox{\Large$v$}}_{D}(x)$ is read as 
\begin{equation}
\overline{\mbox{\Large$v$}}_{D}(x)\equiv
{2(4\pi)^{D/2}\over{\mathrm{tr}}{\bf1}\,\Lambda^{D}}
{\mbox{\Large$v$}}_{D}(m)
={(4\pi)^{D/2}\over{\mathrm{tr}}{\bf1}\,g^{2}\Lambda^{D-2}}
\left(x-{\epsilon\over\Lambda^{2}}\right)
+\int_{1}^{\infty}\!\!d\tau\,
\tau^{-{D\over2}-1}\left\{{\mathrm e}^{-\tau x}
G_{D}(\tau{\cal F})-1\right\}\ .
\label{dlpot}
\end{equation}
The stationary condition, 
$\partial{\mbox{\Large$v$}}_{D}(m)/\partial m=0$, is 
\begin{equation}
{\partial\,\overline{\mbox{\Large$v$}}_{D}(x)\over\partial(m/\Lambda)}
={2m\over\Lambda}
\left[{(4\pi)^{D/2}\over{\mathrm{tr}}{\bf1}\,g^{2}\Lambda^{D-2}}
+f_{D}(x)\right]=0\ ,
\label{statcon}
\end{equation}
where $f_{D}(x)$ is defined as 
\begin{equation}
f_{D}(x)\equiv{\partial\over\partial x}\int_{1}^{\infty}\!\!d\tau\,
\tau^{-{D\over2}-1}{\mathrm e}^{-\tau x}
G_{D}(\tau{\cal F})\ .
\label{fdx}
\end{equation}
Thus the equation for non-trivial solutions of (\ref{statcon}), the gap 
equation, is 
\begin{equation}
-{(4\pi)^{D/2}\over{\mathrm{tr}}{\bf1}\,g^{2}\Lambda^{D-2}}=f_{D}(x)\ .
\label{gapeq}
\end{equation}
Hereafter we designate the stationary point, the solution of the 
gap equation, as $x^{*}(=({m^{*}}^{2}+\epsilon)/\Lambda^{2})$. 
The stability condition 
$\partial^{2}{\mbox{\Large$v$}}_{D}(m)/\partial m^{2}\vert_{m^{*}}\ge0$, 
gives 
\begin{equation}
{\partial^{2}\,\overline{\mbox{\Large$v$}}_{D}(x)\over\partial(m/\Lambda)^{2}}
\Biggr\vert_{x=x^{*}}={4m^{2}\over\Lambda^{2}}f'_{D}(x^{*})\ge0\ ,
\label{stab}
\end{equation}
where use has been made of (\ref{gapeq}), 
and then, the absolute minimum condition, ${\mbox{\Large$v$}}_{D}(m^{*})
-{\mbox{\Large$v$}}_{D}(0)\le0$, leads to 
\begin{eqnarray}
\overline{\mbox{\Large$v$}}_{D}(x^{*})
-\overline{\mbox{\Large$v$}}_{D}(\epsilon/\Lambda^{2})
&=&\int_{\epsilon/\Lambda^{2}}^{x^{*}}\!dx\,f_{D}(x)
+{(4\pi)^{D/2}\over{\mathrm{tr}}{\bf1}\,g^{2}\Lambda^{D-2}}
{{m^{*}}^{2}\over\Lambda^{2}}
\nonumber\\
&=&\int_{\epsilon/\Lambda^{2}}^{x^{*}}dx\,
\left[f_{D}(x)-f_{D}(x^{*})\right]\le0\ .
\label{truemin}
\end{eqnarray}
Finally, we list the integral expression for the effective potential in 
each dimension, 
\begin{eqnarray}
\overline{\mbox{\Large$v$}}_{2}(x)&=&
{2\pi\over g^{2}}\left(x-{\epsilon\over{\Lambda}^2}\right)
+\int_{1}^{\infty}\!\!d\tau\,
\tau^{-2}{\mathrm e}^{-\tau x}
(\tau{\cal F}_{2})\coth(\tau{\cal F}_{2})\ ,
\label{dlp2}\\
\overline{\mbox{\Large$v$}}_{3}(x)&=&
{2{\pi}^{3/2}\over g^{2}\Lambda}\left(x-{\epsilon\over{\Lambda}^2}\right)
+\int_{1}^{\infty}\!\!d\tau\,
\tau^{-{5\over2}}{\mathrm e}^{-\tau x}
(\tau{\cal F}_{3})\coth(\tau{\cal F}_{3})\ ,
\\
\overline{\mbox{\Large$v$}}_{4}(x)&=&
{4{\pi}^{2}\over g^{2}\Lambda^{2}}
\left(x-{\epsilon\over{\Lambda}^2}\right)
+\int_{1}^{\infty}\!\!d\tau\,
\tau^{-3}{\mathrm e}^{-\tau x}
(\tau^{2}{\cal F}_{+}{\cal F}_{-})\coth(\tau{\cal F}_{+})
\coth(\tau{\cal F}_{-})\ ,
\label{dlp4}
\end{eqnarray}
where we have ignored the terms independent of mass and fields. 

\section{Constant Electric Field}\label{electric}

In this section analyses of a constant electric field 
are made. (The case reads, covariantly in the Minkowski 
metric, 
$F_{\mu\nu}F^{\mu\nu}<0$ as well as 
$F_{\mu\nu}\widetilde F^{\mu\nu}=\epsilon_{\mu\nu\lambda\rho}
F^{\mu\nu}F^{\lambda\rho}/2=0$ 
in $4$-dimension.) As was mentioned in the introduction, an imaginary 
part arises in the effective potential: 
the integral in (\ref{fdx}) becomes 
\begin{equation}
\int_{1}^{\infty}\!\!d\tau\,
\tau^{-{D\over2}-1}{\mathrm e}^{-\tau x}
(\tau{{\cal E}})\coth(\tau{{\cal E}})\quad{\mathrm{for}}\quad D=2,3,4\ ,
\label{intE}
\end{equation}
where ${\cal E}$ is the magnitude of the electric field in the 
$\Lambda^{2}$ unit ($0\le{\cal E}<1$). To see the imaginary 
part, go back to the Minkowski space 
via an analytic continuation, ${\cal E}\rightarrow-i{\cal E}$, obtaining 
\begin{eqnarray}
\lefteqn{\int_{1}^{\infty}\!\!d\tau\,
\tau^{-{D\over2}-1}{\mathrm e}^{-\tau x}
(\tau{{\cal E}})\coth(\tau{{\cal E}})}\nonumber\\
&\longrightarrow&
{\mathrm P}\!\int_{1}^{\infty}\!\!d\tau\,
\tau^{-{D\over2}-1}{\mathrm e}^{-\tau x}
(\tau{{\cal E}})\cot(\tau{{\cal E}})
-i\pi\sum_{n=1}^{\infty}\left({\cal E}\over n\pi\right)^{D/2}
{\mathrm e}^{-{n\pi\over {\cal E}}x}
\label{IntMinkowski}
\end{eqnarray}
with ${\mathrm P}$ denoting the principal value. 
The sum of the imaginary part (identical to the modified 
zeta function) is rewritten as, 
\begin{equation}
\pi\sum_{n=1}^{\infty}\left({\cal E}\over n\pi\right)^{D/2}
{\mathrm e}^{-{n\pi\over {\cal E}}x}
={{\cal E}^{D/2}\over\pi^{D/2-1}\Gamma(D/2)}\int_{0}^{\infty}\!\!dt
{t^{D/2-1}\over{\mathrm{e}}^{t+\pi x/{\cal E}}-1}\ .
\label{Imaginaryd}
\end{equation}
In $D=2$, the integral can be performed explicitly, to give 
\begin{equation}
-{\cal E}\,{\mathrm{ln}}\!\left(1-{\mathrm e}^{-{\pi x\over{\cal E}}}\right)\ .
\label{Imaginary2d}
\end{equation}

\begin{figure}[ht]
\centering\leavevmode
{\epsfysize=6cm\epsfbox{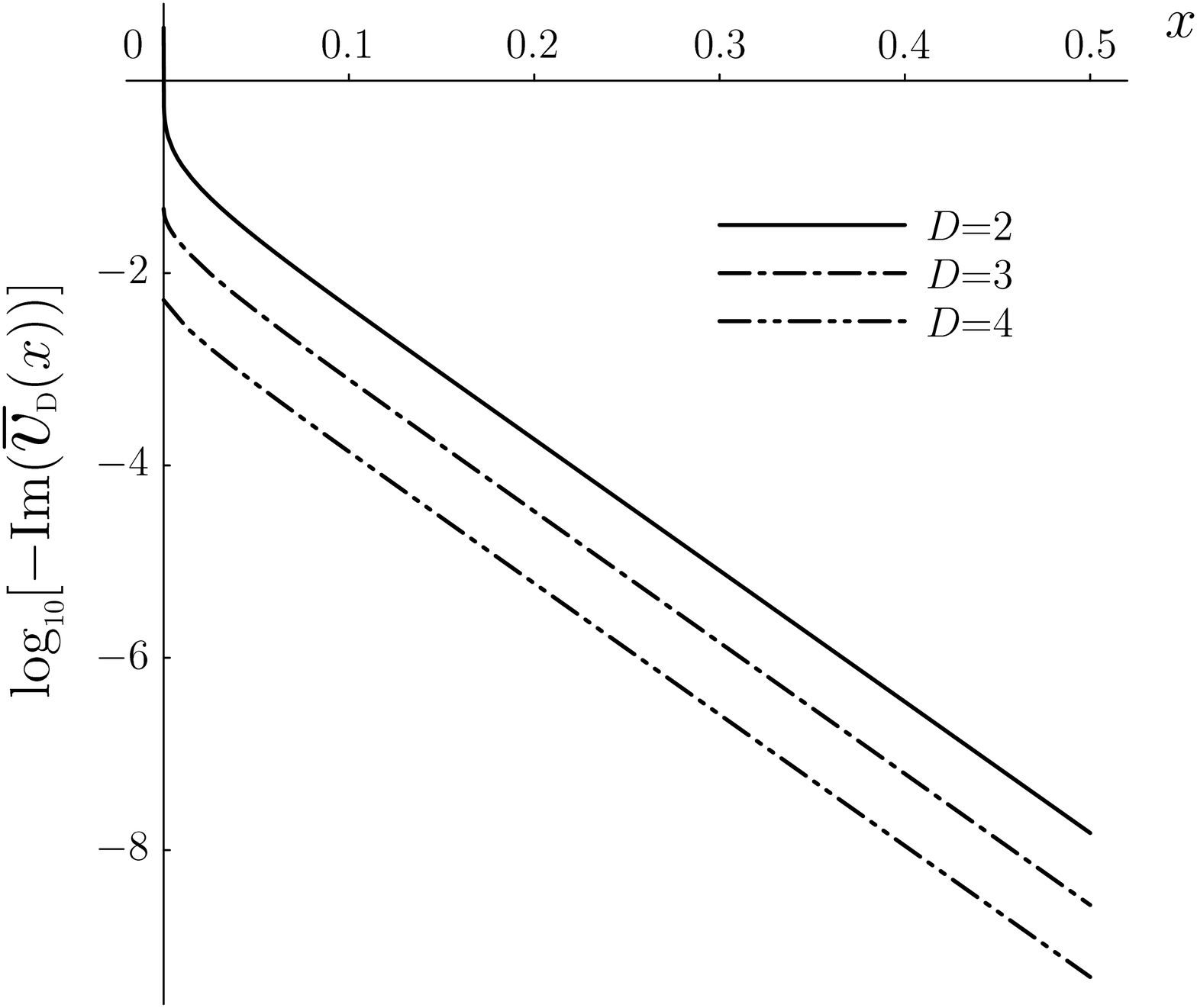}}
\caption{Behavior of the imaginary part of the potential: we write the 
quantity of (\ref{Imaginaryd}) as 
$-{\mathrm Im}\overline{\mbox{\Large$v$}}_{D}(x)$ and put ${\cal E}=0.1$. 
The imaginary part is exponentially dumping as $x$ goes larger and is 
smaller than $10^{-2}$ when $x>{\cal E}$. 
${\mathrm Im}\overline{\mbox{\Large$v$}}_{D}(0)$ 
is finite in $D=3$ and $D=4$ but divergent in $D=2$.}
\label{imaginarypart}
\end{figure}

In view of (\ref{Imaginaryd}) (or (\ref{Imaginary2d})), the imaginary 
part becomes significant when ${\cal E}>x$ ($E>m^{2}$), which is a 
rather well-known result from a one-body problem of the Dirac 
particle under an external potential --- the Klein paradox. 
The appearance of the imaginary part implies the vacuum instability by 
means of external electric fields \cite{GMR}. The initial vacuum goes 
to a new 
one with emissions of particle pairs to neutralize the electric field 
which, however, has assumed constant so that the process will never end. 
To avoid the situation electric fields must be localized \cite{ST} or 
the condition, 
\begin{equation}
{\cal E}<x\ ,
\label{Ex}
\end{equation}
must be assumed. (See Fig.~\ref{imaginarypart}.) Moreover 
if a perturbative expansion in terms of the 
electric field is employed we cannot see any imaginary part at all 
since ${\cal E}=0$ is an essential singularity. 
The program matches with our strategy. As was stated in the 
introduction, our 
interest lies in seeking a change when external fields are present, 
then it is enough to regard external fields to be very small: 
perturbations would be useful. 

Here we should make three comments.  
\begin{itemize}
\item[(i)] According to the condition (\ref{Ex}) there is no need for 
an infrared cut-off in this case so that we put $\epsilon=0$ hereafter. 
\item[(ii)] It is impossible to talk about criticality of electric 
fields \cite{KL}, since it is defined through a massless condition under the 
variation of electric fields, which contradicts the condition (\ref{Ex}). 
\item[(iii)] The calculation is performed in the Euclidean world 
throughout. (In the above, we visited in the Minkowski space just 
because of glancing at the infrared structure.)
\end{itemize}

With these spirits we first expand the integrand in (\ref{intE}) 
such that
\begin{equation}
t\coth t=1+{t^{2}\over3}+O(t^{4})
\label{expansionE}
\end{equation}
then integrate each term, and perform the analytic continuation 
${\cal E}\rightarrow -i{\cal E}$, 
to obtain \\
$D=2$ case: 
\begin{equation}
\overline{\mbox{\Large$v$}}_{2}(x)={2\pi\over g^{2}}x+xE_{i}(-x)
+\left(1-{{\cal E}^{2}\over3x}\right){\mathrm e}^{-x}-1\ ,
\label{pot2dE}
\end{equation}
\begin{equation}
-{2\pi\over g^{2}}=f_{2}(x)=Ei(-x)+{{\cal E}^{2}\over3}\left(
{1\over x}+{1\over x^{2}}\right){\mathrm e}^{-x}\ ,
\label{fx2E}
\end{equation}
$D=3$ case: 
\begin{equation}
\overline{\mbox{\Large$v$}}_{3}(x)={2\pi^{3/2}\over g^{2}\Lambda}x
+{1\over3}\left(4x-{{\cal E}^{2}\over x}\right)x^{1/2}{\mit\Gamma}(1/2,x)
+{2\over3}(1-2x){\mathrm e}^{-x}-{2\over3}\ ,
\end{equation}
\begin{equation}
-{2\pi^{3/2}\over g^{2}\Lambda}=f_{3}(x)
=\left(2+{{\cal E}^{2}\over6x^{2}}\right)x^{1/2}{\mit\Gamma}(1/2,x)
-\left(2-{{\cal E}^{2}\over3x}\right){\mathrm e}^{-x}\ ,
\end{equation}
$D=4$ case: 
\begin{equation}
\overline{\mbox{\Large$v$}}_{4}(x)={4\pi^{2}\over g^{2}\Lambda^{2}}x
-\left({x^{2}\over2}-{{\cal E}^{2}\over3}\right)E_{i}(-x)
+{1\over2}(1-x){\mathrm e}^{-x}-{1\over2}\ ,
\end{equation}
\begin{equation}
-{4\pi^{2}\over g^{2}\Lambda^{2}}=f_{4}(x)=-xE_{i}(-x)
-\left(1-{{\cal E}^{2}\over3x}\right){\mathrm e}^{-x}\ .
\label{gap4dE}
\end{equation}
Here use has been made of  
\begin{equation}
-E_{i}(-z)=\int_{z}^{\infty}\!\!dt\,{\mathrm e}^{-t}t^{-1}
=\int_{1}^{\infty}\!\!dt{\mathrm e}^{-tz}t^{-1}\ ,
\label{ei}
\end{equation}
and 
\begin{equation}
{\mit\Gamma}(1/2,z)=\int_{z}^{\infty}\!\!dt\,{\mathrm e}^{-t}t^{-1/2}\ .
\label{igamma}
\end{equation}
As was expected from the above argument there is no 
imaginary part. It should be noted that our perturbation expansion 
(\ref{expansionE}) is of course an asymptotic expansion to the 
effective potential (\ref{dlp2}) -- (\ref{dlp4}) since we have 
regarded $t$ in (\ref{expansionE}) as $\tau{\cal E}$ which apparently 
exceeds the radius of convergence when $\tau\rightarrow\infty$. 

\begin{figure}[ht]
\centering
(a)
{\epsfysize=6cm\epsfbox{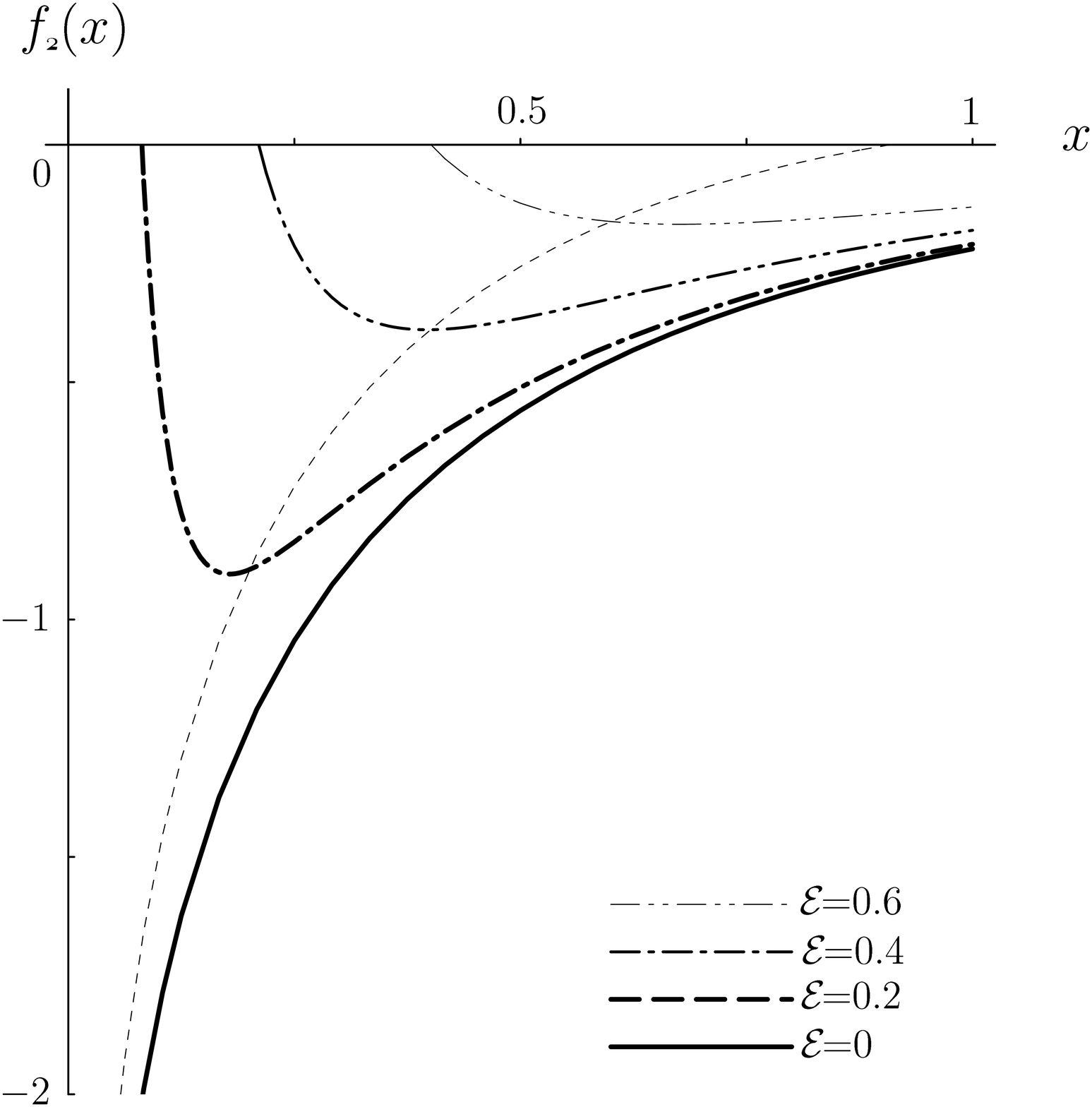}}
(b)
{\epsfysize=6cm\epsfbox{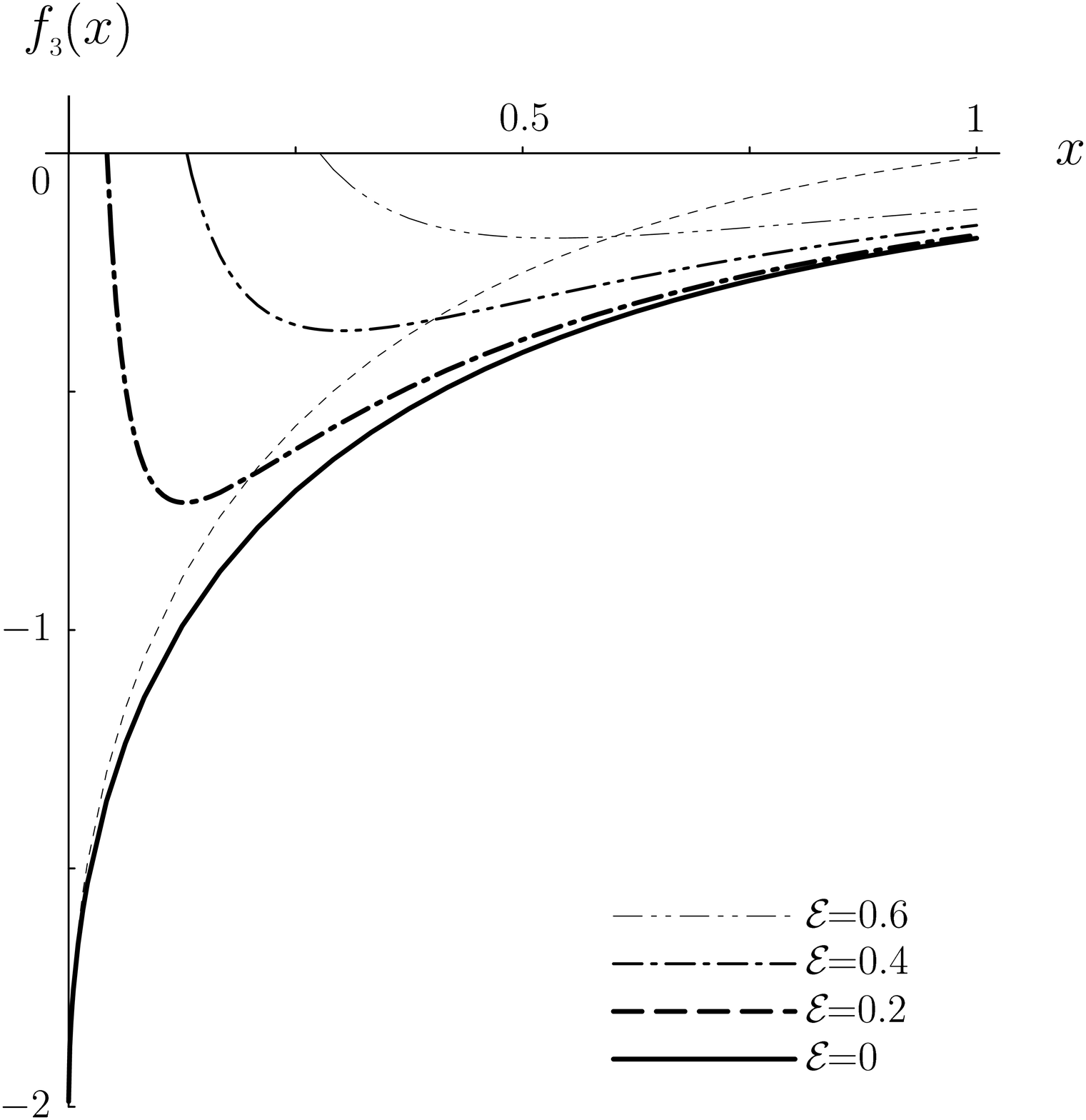}}\\
(c)
{\epsfysize=6cm\epsfbox{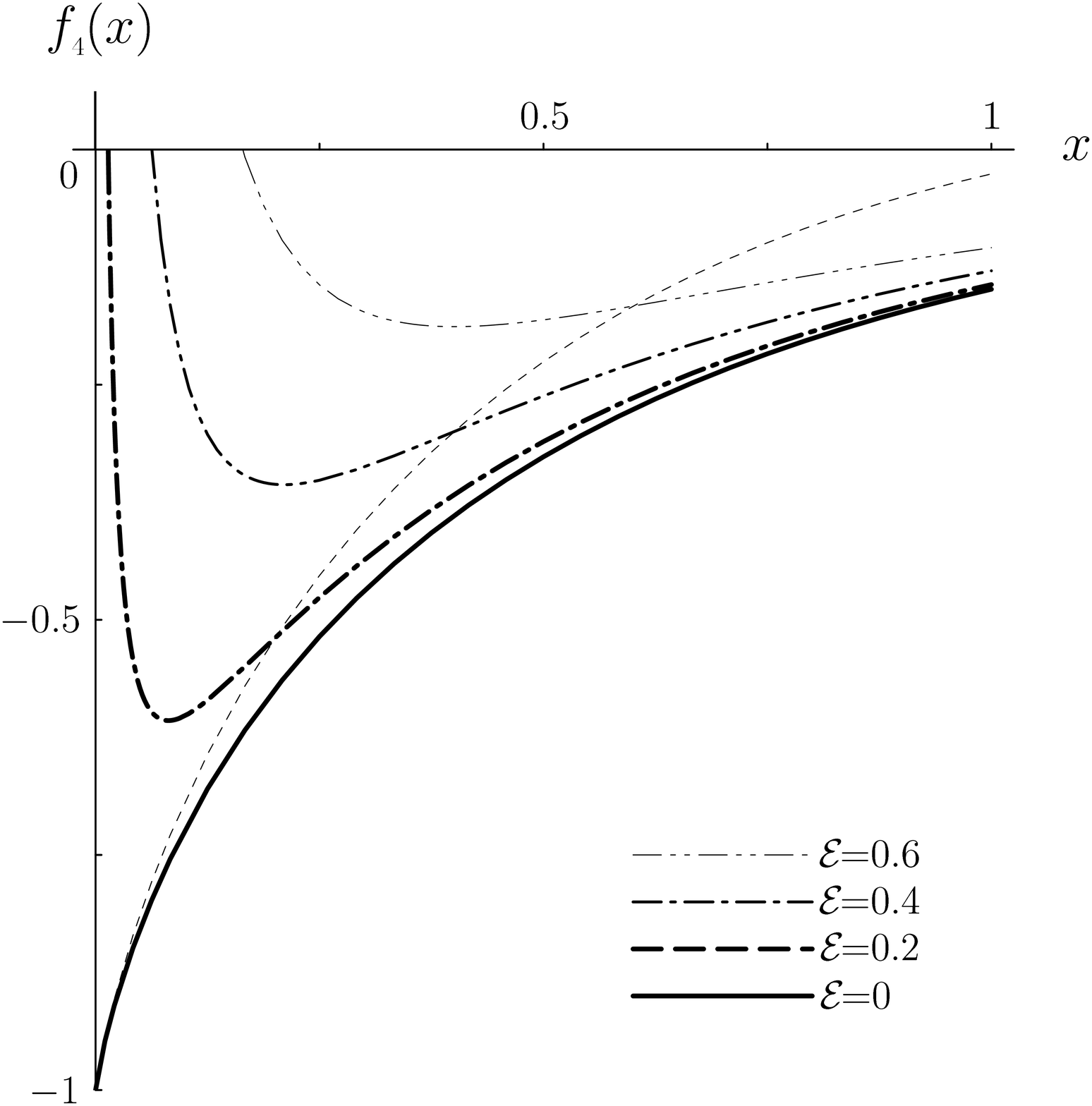}}
\caption{$f_{D}(x)$ for a pure electric field in (a) 
$D=2$, (b) $D=3$, and (c) $D=4$ for several fixed values of ${\cal E}$. 
In each plot, the thin-dashed line 
denotes $f_{D}(x)\vert_{{\cal E}=x}$ and shows the lower bound of $x$ 
(\ref{Ex}): the vacuum is unstable in the left region to 
the line.} 
\label{fxE}
\end{figure}

\begin{figure}[ht]
\centering
(a)
{\epsfysize=6cm\epsfbox{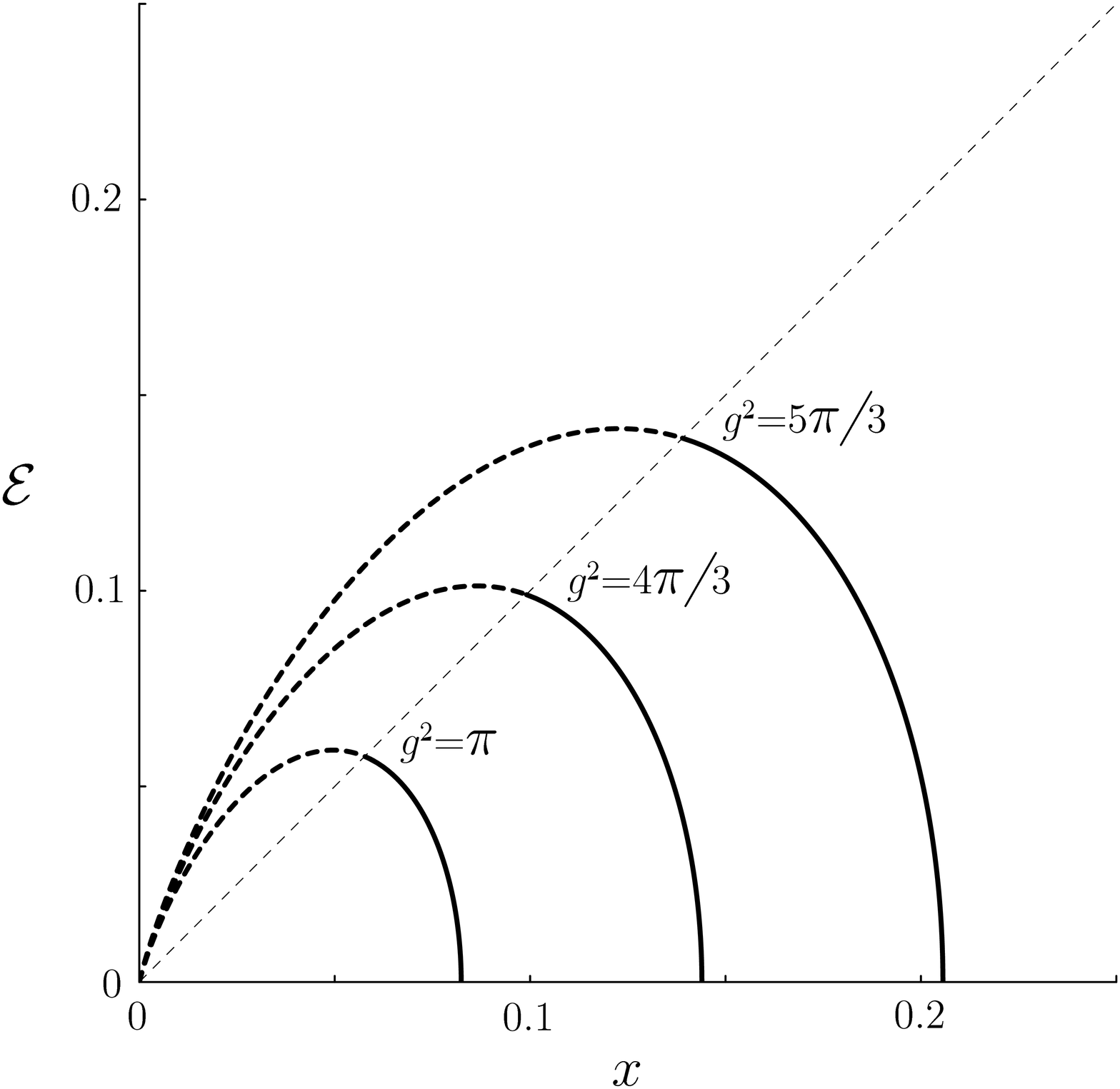}}
(b)
{\epsfysize=6cm\epsfbox{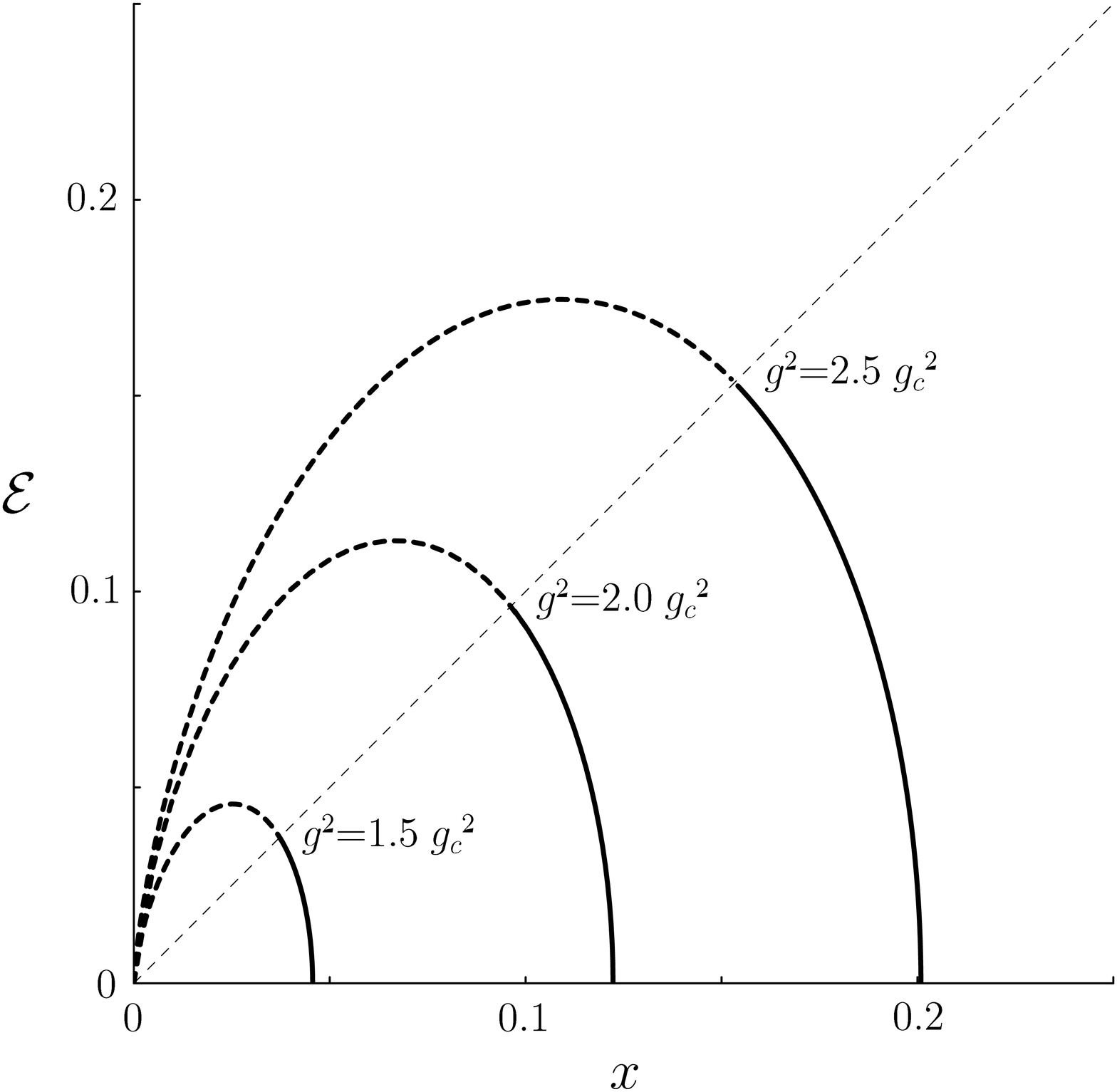}}\\
(c)
{\epsfysize=6cm\epsfbox{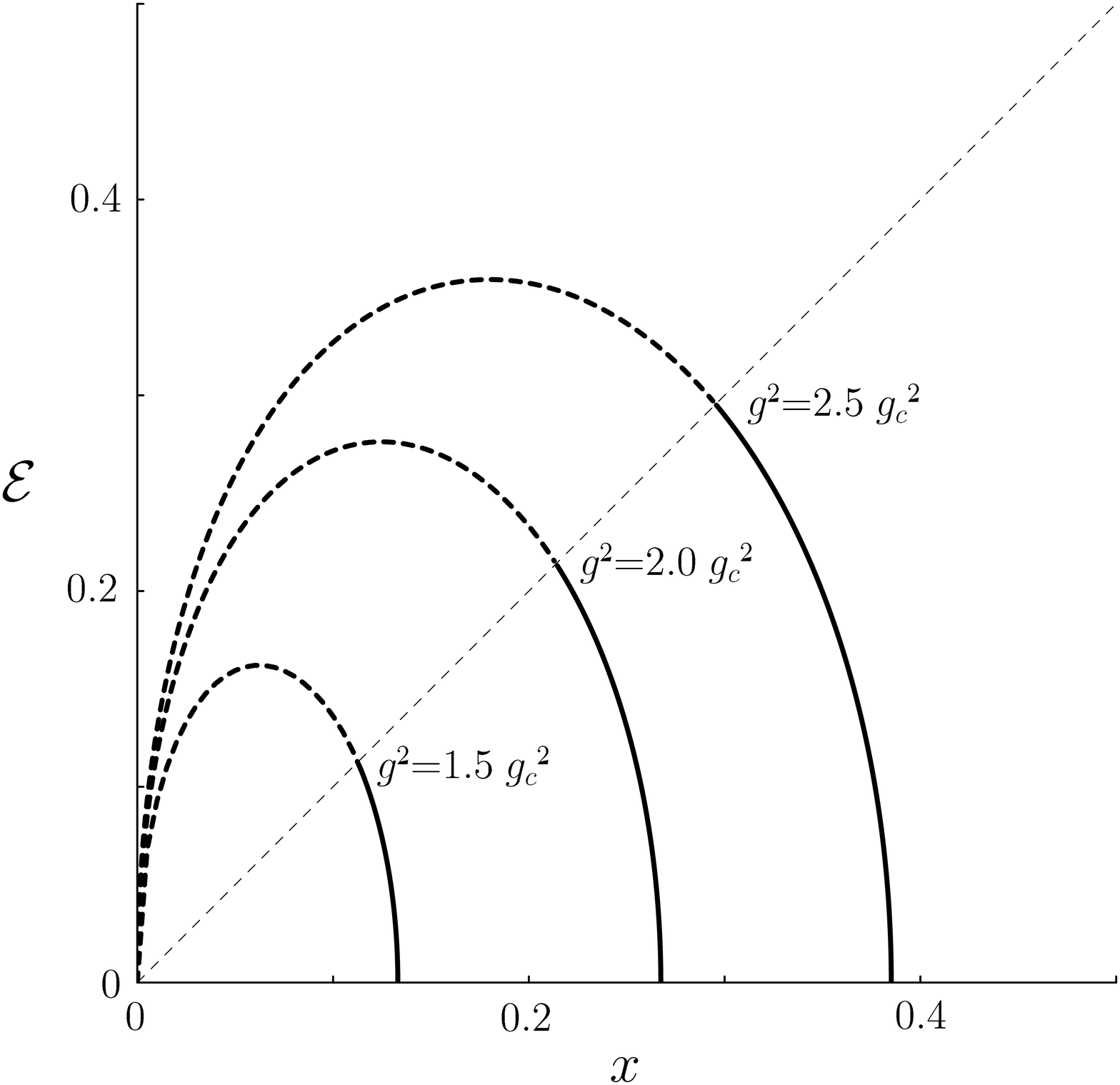}}
\caption{Relations between the magnitude of the electric field 
${\cal E}$ and the dynamically generated mass $x$ for several fixed 
values of $g^{2}$ in (a) $D=2$, (b) $D=3$, and (c) $D=4$. In each plot, 
the thin-dashed 
line shows the lower bound of $x$ (\ref{Ex}). $g_{c}^{2}$ 
in (b) and (c) denote the critical couplings 
without external fields in each dimension and are given by 
$\pi^{3/2}/\Lambda$ and $4\pi^{2}/\Lambda^{2}$ respectively. This 
shows that the mass becomes smaller when the electric field goes larger. 
}
\label{phaseE}
\end{figure}

We plot $f_{D}(x)$ in Fig.~\ref{fxE} (a)--(c) for 
several fixed values of ${\cal E}$. First note that in each plot, 
the thin-dashed line denoting 
$f_{D}(x)\vert_{{\cal E}=x}$ shows the lower bound of $x$ (\ref{Ex}): 
the vacuum is unstable in the left region to the line. 
From the figures we can see that the 
external electric field makes the mass smaller 
for a given coupling. This can more 
easily be seen from Fig.~\ref{phaseE} (a)--(c) where the relation between 
${\cal E}$ and $x$ 
are depicted for several fixed values of $g^{2}$. In each plot, 
the thin-dashed line denoting ${\cal E}=x$, again shows the lower 
bound of $x$ (\ref{Ex}); the vacuum is unstable in the upper triangular 
region. $g_{c}^{2}$ in (b) and (c) are the critical couplings 
without any external fields in $D=3$ and $4$ and given by 
$\pi^{3/2}/\Lambda$ and $4\pi^{2}/\Lambda^{2}$ respectively. 

\begin{figure}[ht]
\centering\leavevmode
{\epsfysize=6cm\epsfbox{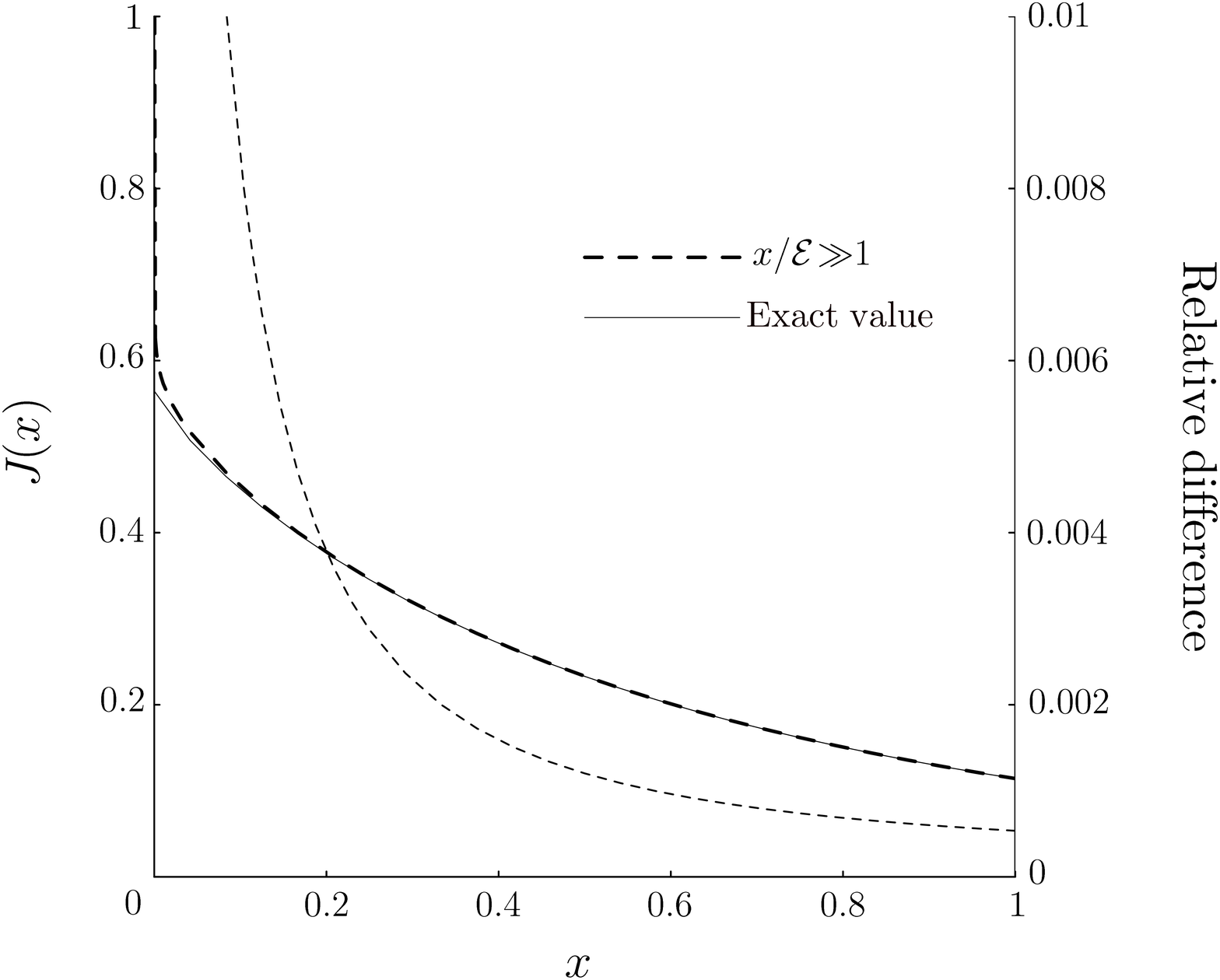}}
\caption{Plausibility of the asymptotic expansion in the pure 
electric field case: we write the integral (\ref{intE}) in $D=4$ as 
$J(x)\equiv{\cal E}\int_{1}^{\infty}\!\!d\tau\,\tau^{-2}
{\mathrm e}^{-\tau x}\coth(\tau{\cal E})$ and put ${\cal E}=0.25$. The 
thin line denotes the (numerically evaluated) exact value and the 
dashed line represents the asymptotic expansion up to the second 
term. The dotted line denotes the relative difference 
$[(Approx.)-(Exact)]/(Exact)$ using right-hand scale. 
Matching is excellent for any $x$ down to ${\cal E}$.}
\label{plauseE}
\end{figure}

Finally it is necessary to discuss plausibility of our approximation. 
By changing the integration variable such as $\tau x\rightarrow\tau$, 
(\ref{intE}) is rewritten as 
\begin{equation}
x^{{D\over2}}\int_{x}^{\infty}\!\!d\tau\,
\tau^{-{D\over2}-1}{\mathrm e}^{-\tau}
({\tau{\cal E}\over x})\coth({\tau{\cal E}\over x})\ .
\label{intEdash}
\end{equation}
Therefore our approximation becomes good when 
\begin{equation}
{x\over{\cal E}}\rightarrow\infty\ .
\end{equation}
However, as is seen from Fig.~\ref{plauseE}, where the integral 
(\ref{intE}) for $D=4$ with ${\cal E}=0.25$ is depicted as the function 
of $x$, matching is excellent for any $x$ down to ${\cal E}$. 

\section{Constant Magnetic Field}\label{magnetic}

\begin{figure}[ht]
\centering
(a)
{\epsfysize=6cm\epsfbox{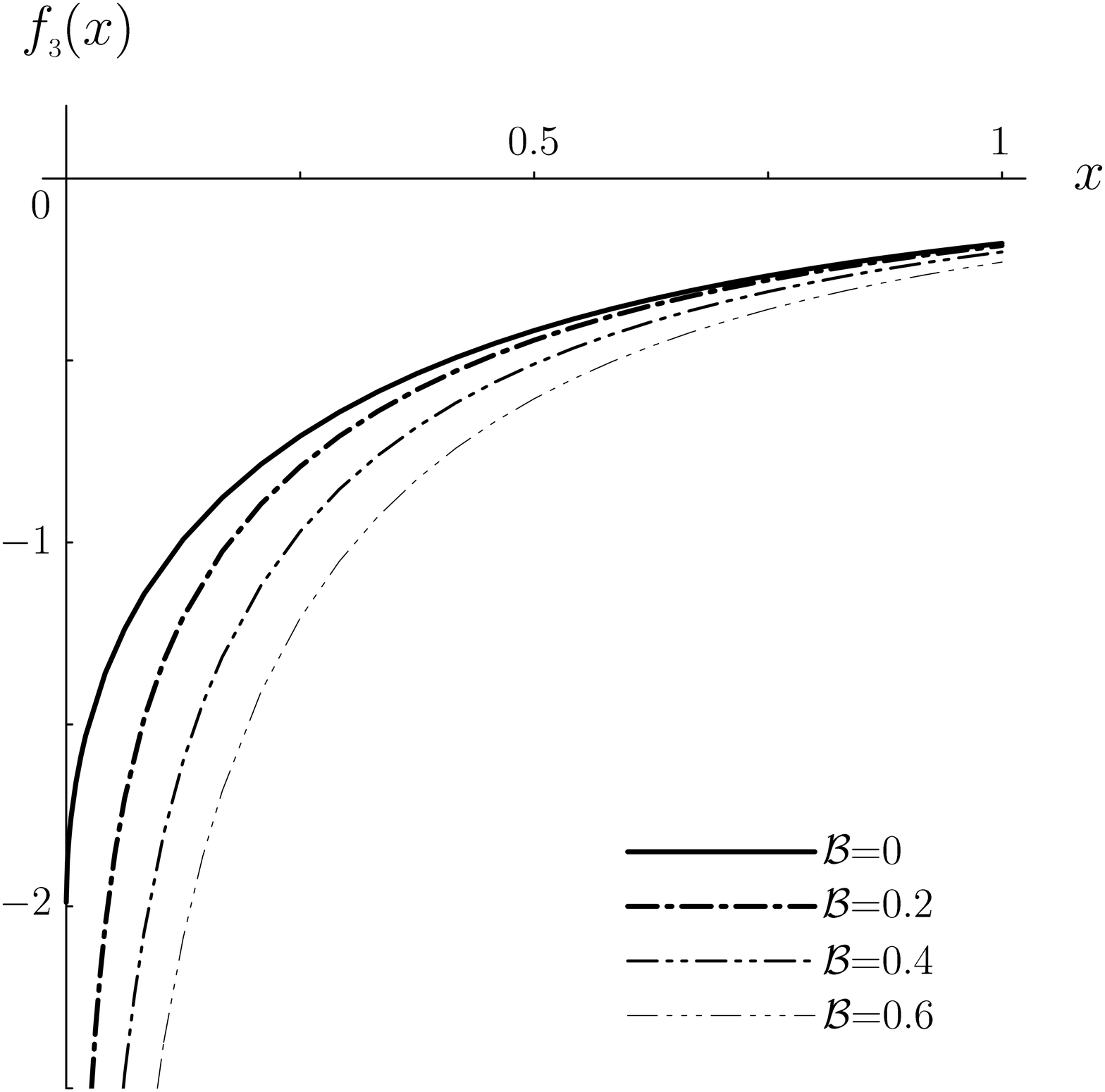}}
(b)
{\epsfysize=6cm\epsfbox{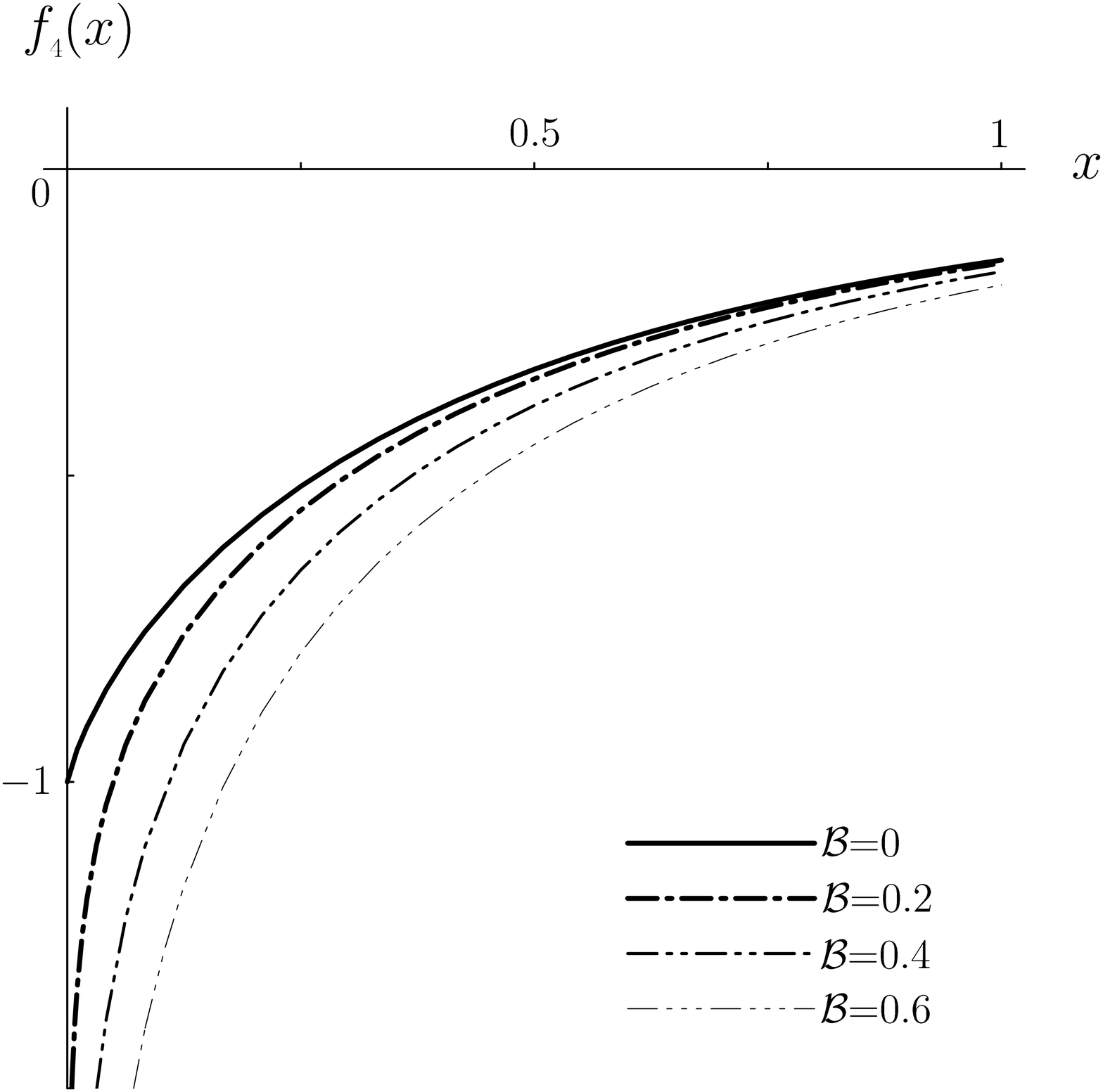}}
\caption{$f_{D}(x)$ for a pure magnetic field in 
(a) $D=3$ and (b) $D=4$ for several fixed values of ${\cal B}$. 
The minimum of $x$ is 
$\epsilon/\Lambda^{2}$ so that there exists a finite critical 
coupling as far as $\epsilon\ne0$. However when $\epsilon\rightarrow0$ 
critical couplings go to zero ($f_{D}(x)\rightarrow-\infty$) for any 
non-zero ${\cal B}$. }
\label{fxM}
\end{figure}

The case for a constant magnetic field is considered in this 
section. (This reads, covariantly in the Minkowski space, 
$F_{\mu\nu}F^{\mu\nu}>0$ with 
$F_{\mu\nu}\widetilde F^{\mu\nu}=0$ in $4$-dimension.) 
Contrary to the previous case, 
there arises no 
imaginary part in the effective potential: 
the integral in (\ref{fdx}) reads 
\begin{equation}
\int_{1}^{\infty}\!\!d\tau\,
\tau^{-{D\over2}-1}{\mathrm e}^{-\tau x}
\tau{{\cal B}}\coth(\tau{{\cal B}})
\label{IntMinkowskiB}
\end{equation}
with ${\cal B}$ denoting a magnitude of the magnetic field in the 
$\Lambda^{2}$ unit ($0\le{\cal B}<1$). 
There is no need 
for an analytic continuation so that any imaginary part does not occur. 
Therefore, unlike the electric case, there is 
no restriction on ${\cal B}$ and $x$ (except for $0\le{\cal B}<1$ and 
$0\le x<1$) such as (\ref{Ex}), which enforce us to keep 
$\epsilon$ non zero and to modify the na\"{\i}ve expansion 
(\ref{expansionE}), since in this case ${\cal B}/x$ can become large 
contrary to the electric case where ${\cal E}/x<1$ from (\ref{Ex}). 
We thus arrange (\ref{IntMinkowskiB}) such that 
\begin{eqnarray}
\lefteqn{\int_{1}^{\infty}\!\!d\tau\,\tau^{-{D\over2}-1}
{\mathrm e}^{-\tau x}
(\tau{\cal B})\coth(\tau{\cal B})}
\nonumber\\
&=&{\cal B}\int_{1}^{\infty}\!\!d\tau\,\tau^{-{D\over2}}{\mathrm e}^{-\tau x}
+\int_{1}^{\infty}\!\!d\tau\,\tau^{-{D\over2}-1}
{\mathrm e}^{-\tau(x+2{\cal B})}
{2\tau{\cal B}\over1-{\mathrm e}^{-2\tau{\cal B}}}\ .
\label{IntMinkowskiB2}
\end{eqnarray}
The last factor in the second integral can be expanded as 
\begin{equation}
{t\over{\mathrm e}^{t}-1}=1-{t\over2}+{t^{2}\over12}+O(t^{4})\ ,
\label{expansion}
\end{equation}
and then all the term can be 
evaluated with the use of (\ref{ei}) or (\ref{igamma}). This gives 
an improved asymptotic expansion. The results are read as follows: \\
$D=3$ case: 
\begin{eqnarray}
\overline{\mbox{\Large$v$}}_{3}(x)&=&{2\pi^{3/2}\over g^{2}\Lambda}
\left(x-{\epsilon\over\Lambda^{2}}\right)
+{1\over3}\left[2(2x+{\cal B})+{{\cal B}^{2}\over x+2{\cal B}}\right]
(x+2{\cal B})^{1/2}{\mit\Gamma}(1/2,x+2{\cal B})
\nonumber\\
&&+{2\over3}(1-2x-{\cal B}){\mathrm e}^{-(x+2{\cal B})}-2{\cal B}
\left[x^{1/2}{\mit\Gamma}(1/2,x)-{\mathrm e}^{-x}\right]-{2\over3}\ ,
\end{eqnarray}
\begin{eqnarray}
-{2\pi^{3/2}\over g^{2}\Lambda}=f_{3}(x)&=&\left[
2-{{\cal B}\over x+2{\cal B}}-{{\cal B}^{2}\over6(x+2{\cal B})^{2}}\right]
(x+2{\cal B})^{1/2}{\mit\Gamma}(1/2,x+2{\cal B})
\nonumber\\
&&-\left[2+{{\cal B}^{2}\over3(x+2{\cal B})}\right]
{\mathrm e}^{-(x+2{\cal B})}-{\cal B}x^{-1/2}{\mit\Gamma}(1/2,x)\ .
\label{gapm3}
\end{eqnarray}
$D=4$ case: 
\begin{eqnarray}
\overline{\mbox{\Large$v$}}_{4}(x)&=&{4\pi^{2}\over g^{2}\Lambda^{2}}
\left(x-{\epsilon\over\Lambda^{2}}\right)
-\left({x^{2}\over2}+x{\cal B}+{{\cal B}^{2}\over3}\right)
E_{i}(-(x+2{\cal B}))
\nonumber\\
&&+{1\over2}(1-x){\mathrm e}^{-(x+2{\cal B})}
+{\cal B}\left[xE_{i}(-x)+{\mathrm e}^{-x}\right]-{1\over2}\ ,
\end{eqnarray}
\begin{equation}
-{4\pi^{2}\over g^{2}\Lambda^{2}}=f_{4}(x)=
-(x+{\cal B})E_{i}(-(x+2{\cal B}))
-\left(1+{{\cal B}^{2}\over3(x+2{\cal B})}\right)
{\mathrm e}^{-(x+2{\cal B})}+{\cal B}E_{i}(-x)\ .
\label{gapm4}
\end{equation}
We plot $f_{D}(x)$ in Fig.~\ref{fxM} (a) and (b) for 
several fixed values of ${\cal B}$, from which we see 
that any point on the line fulfills the condition 
$f'_{D}(x^{*})\ge0$, (\ref{stab}). The 
condition (\ref{truemin}) is also satisfied, since (\ref{truemin}) 
is always true for any monotonically increasing $f_{D}(x)$. Since the 
minimum of $x$ is $\epsilon/\Lambda^{2}$, a finite critical coupling exists 
as a solution of (\ref{gapeq}) at $x=\epsilon/\Lambda^{2}$; 
\begin{equation}
g^{2}_{c}(\epsilon)\equiv-{(4\pi)^{D/2}
\over{\mathrm{tr}}{\bf1}\,\Lambda^{D-2}f_{D}(\epsilon/\Lambda^{2})}
\end{equation}
for a fixed ${\cal B}$. However, when 
$\epsilon/\Lambda^{2}\rightarrow0$, $f_{D}(\epsilon/\Lambda^{2})$ 
behaves as 
\begin{eqnarray}
f_{3}(\epsilon/\Lambda^{2})&\sim&-{\cal B}(\epsilon/\Lambda^{2})^{-1/2}
\Gamma(1/2,\epsilon/\Lambda^{2})
+O\left((\epsilon/\Lambda^{2})\right)\ ,
\\
f_{4}(\epsilon/\Lambda^{2})&\sim&{\cal B}E_{i}(-\epsilon/\Lambda^{2})
+O\left((\epsilon/\Lambda^{2})\right)\ ,
\label{f4x0}
\end{eqnarray}
so that $f_{D}(\epsilon/\Lambda^{2})\rightarrow-\infty$, 
as far as ${\cal B}\ne0$. This implies that 
{\it the critical coupling goes to zero, $g_{c}\rightarrow0$, for 
any non-zero ${\cal B}$}. The situation is similar to the 2-dimensional 
case without external fields, where from (\ref{fx2E}) 
\begin{equation}
f_{2}(x)=E_{i}(-x)\ ,
\end{equation}
which is again divergent when $x\rightarrow0$. 
Keeping IR cut-off $\epsilon$ finite, 
we obtain a finite critical coupling 
\begin{equation}
g_{c}^{2}(\epsilon)=-{2\pi\over f_{2}(\epsilon/\Lambda^{2})}\ ,
\end{equation}
which also becomes zero when $\epsilon/\Lambda^{2}\rightarrow0$. 
(Compare Fig.~\ref{fxM} (b) with the ${\cal E}=0$ case of 
Fig.~\ref{fxE} (a).) 
Gusynin, Miransky, and Shovkovy have 
interpreted this similarity in terms of the dimensional reduction 
\cite{GMS2}. 

\begin{figure}[ht]
\centering
(a)
{\epsfysize=6cm\epsfbox{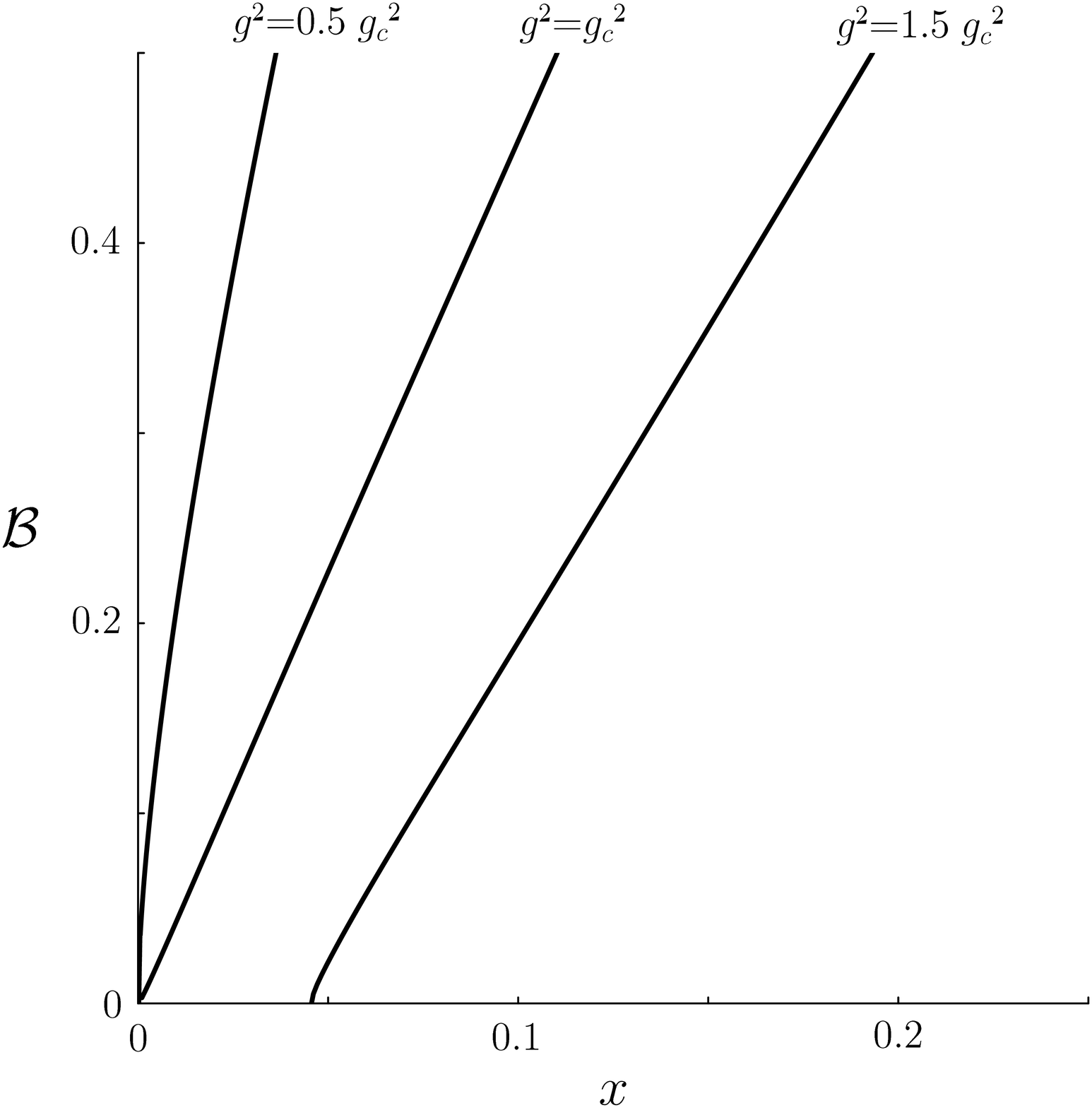}}
(b)
{\epsfysize=6cm\epsfbox{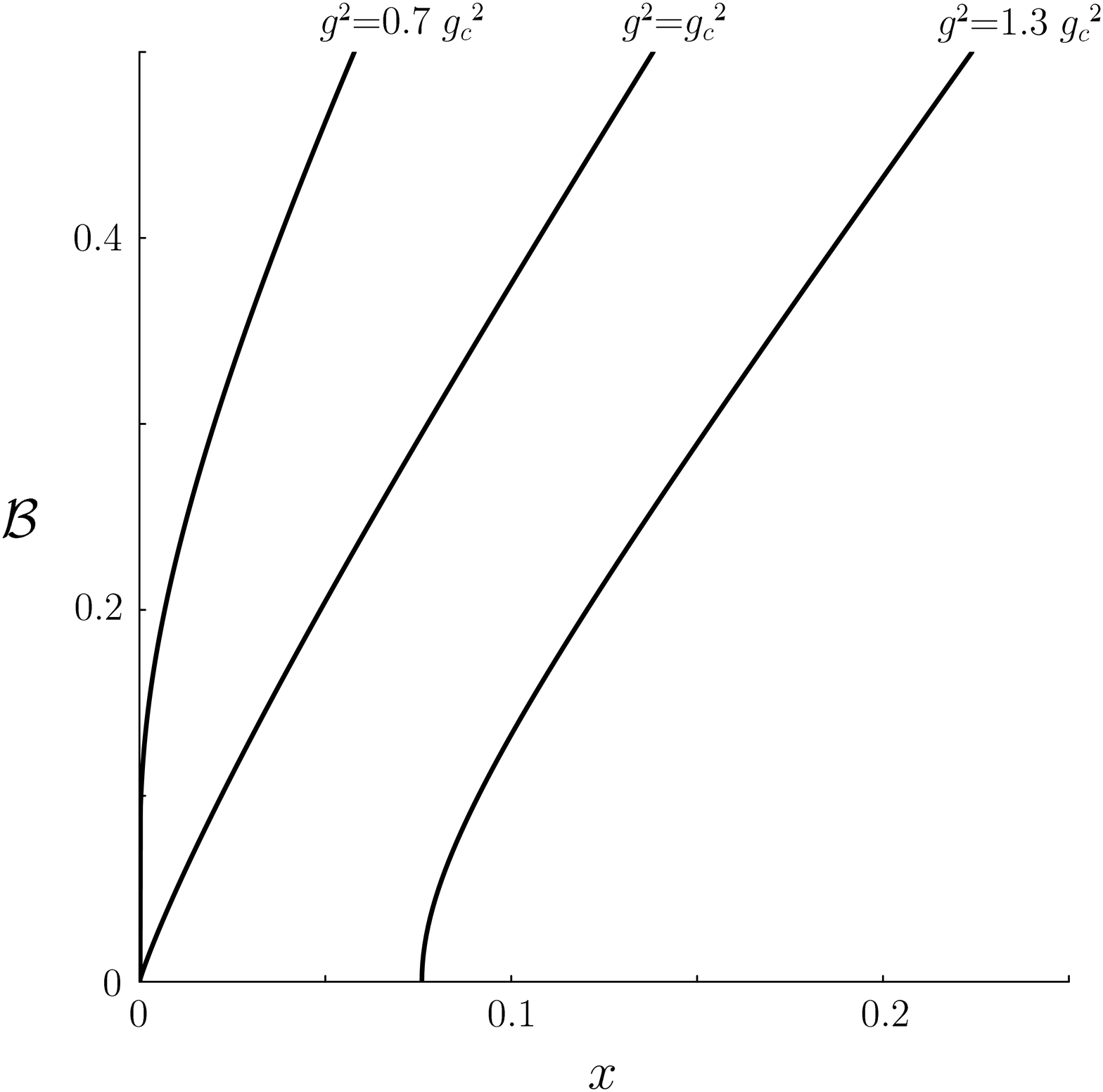}}
\caption{Relations between the magnitude of the magnetic field 
${\cal B}$ and the dynamically generated mass $x$ for several fixed 
values of $g^{2}$ in (a) $D=3$ and (b) $D=4$. 
$g_{c}^{2}$ denotes the critical couplings without external fields 
given by $\pi^{3/2}/\Lambda$ in 3-dimension and $4\pi^{2}/\Lambda^{2}$ 
in 4-dimension. This shows that a magnetic field prompts the 
generation of mass thus enhances $\chi$SB. }
\label{phaseM}
\end{figure}

We can also see, from Fig.~\ref{fxM} (a) and (b), 
that the dynamically generated 
mass in a constant magnetic field is larger than that 
without any external field for a fixed coupling. This can more easily be 
seen from Fig.~\ref{phaseM} (a) and (b), where the relations between 
${\cal B}$ and $x$ is depicted for 
several fixed values of $g^{2}$. 

\begin{figure}[ht]
\centering\leavevmode
{\epsfysize=6cm\epsfbox{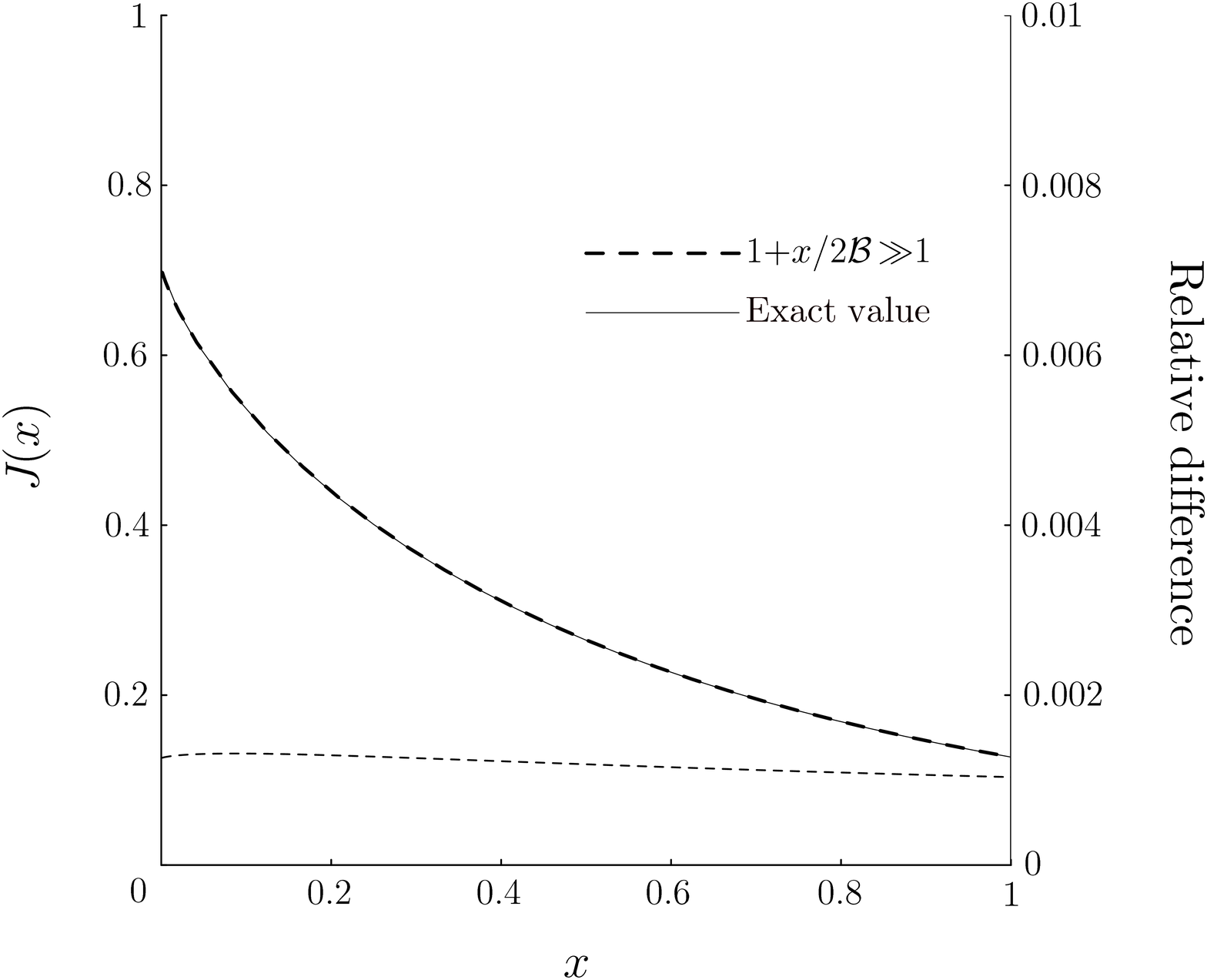}}
\caption{
Plausibility of the improved asymptotic expansion in the pure 
magnetic field case: we write the integral (\ref{IntMinkowskiB}) 
in $D=4$ as 
$J(x)\equiv{\cal B}\int_{1}^{\infty}\!\!d\tau\,\tau^{-2}
{\mathrm e}^{-\tau x}\coth(\tau{\cal B})$ and put 
${\cal B}=0.5$. The thin line denotes 
the (numerically evaluated) exact value and the dashed line 
represents the improved asymptotic expansion up to the first three 
terms. The dotted line denotes the relative difference 
$[(Approx.)-(Exact)]/(Exact)$. 
Matching is significantly excellent for all values of $x$. 
}
\label{plauseB}
\end{figure}

We adopt the improved expansion (\ref{IntMinkowskiB2}) with 
(\ref{expansion}) against the previous one (\ref{expansionE}). The 
expansion becomes exact when 
$1+x/(2{\cal B})\rightarrow\infty$, which can be recognized by 
looking at the second term in the 
right-hand side of (\ref{IntMinkowskiB2}). However as is seen from 
Fig.~\ref{plauseB}, 
matching to the exact value is much more excellent than the 
previous case in Fig.~\ref{plauseE}, even if 
only the first three terms are taken into account. 

\section{General Constant Fields in D=4}\label{electromagnetic}

So far, in D=4, the case $F_{\mu\nu}\tilde{F}^{\mu\nu}=0$ (in the 
Minkowski metric), that is, 
${\bbox{B\cdot E}}=0$ has been assumed, but in this section a more 
general case $F_{\mu\nu}\tilde{F}^{\mu\nu}\neq 0$, is considered. 
We vary $F_{\mu\nu}\tilde{F}^{\mu\nu}$ with $F_{\mu\nu}{F}^{\mu\nu}$ 
being fixed, which is interpreted, for example, such that 
the angle between 
$\bbox{B}$ and $\bbox{E}$ is shifted from $\pi/2$, 
while keeping the magnitude of $\bbox{B}$ and $\bbox{E}$ fixed. 

In this case, imaginary parts in the effective potential are also 
unavoidable. To see this, the analytic continuation 
$\vert{\bbox{E}}\vert\rightarrow-i\vert{\bbox{E}}\vert$ is performed 
in the third relation in (\ref{Fs}) and a covariant notation is 
adopted, such that 
\begin{eqnarray}
F_{\mu\nu}F^{\mu\nu}&=&
2({\bbox{B}}^{2}-{\bbox{E}}^{2})\ ,
\nonumber\\
F_{\mu\nu}\widetilde F^{\mu\nu}&=&-4{\bbox{B\cdot E}}\ ,
\end{eqnarray}
giving 
\begin{eqnarray}
{F}_{+}^{E}&\rightarrow&
{1\over2}\sqrt{\sqrt{(F_{\mu\nu}F^{\mu\nu})^{2}
+(F_{\mu\nu}\widetilde F^{\mu\nu})^2}+F_{\mu\nu}F^{\mu\nu}}
\equiv F_{+}\ , \\
{F}_{-}^{E}&\rightarrow&
{i\over2}{\mathrm{sgn}}(F_{\mu\nu}\widetilde F^{\mu\nu})
\sqrt{\sqrt{(F_{\mu\nu}F^{\mu\nu})^{2}
+(F_{\mu\nu}\widetilde F^{\mu\nu})^2}-F_{\mu\nu}F^{\mu\nu}}
\equiv-i{F}_{-}\ ,
\label{Fpm}
\end{eqnarray}
where ${\mathrm{sgn}}(x)$ designates the sign function 
\begin{equation}
{\mathrm{sgn}}(x)\equiv\left\{
\begin{array}{rl}
1&(x\ge0)\ ,\\  
-1&(x<0)\ ,
\end{array}\right.
\end{equation}
and the notation $F_{\pm}^{E}$ has been employed to distinguish the 
Euclidean quantities from the Minkowski ones. 
Since there exists a suitable Lorentz frame where 
$\vert{\bbox B}'\vert=F_{+}$, 
$\vert{\bbox E}'\vert=\vert F_{-}\vert$, and 
${\bbox B}'\parallel{\bbox E}'$, we regard $F_+$ ($F_-$) 
as a Lorentz invariant magnetic (electric) field. 
The integral in (\ref{dlp4}) becomes, after the analytic continuation, to
\begin{eqnarray}
\lefteqn{\int_{1}^{\infty}\!\!d\tau\,\tau^{-3}{\mathrm e}^{-\tau x}
(\tau^{2}{\cal F}_{+}^{E}{\cal F}_{-}^{E})\coth(\tau{\cal F}_{+}^{E})
\coth(\tau{\cal F}_{-}^{E})}   \nonumber  \\
&\longrightarrow&{\mathrm P}\int_{1}^{\infty}\!\!d\tau\,\tau^{-3}
{\mathrm e}^{-\tau x}(\tau^{2}{\cal F}_{+}{\cal F}_{-})
\coth(\tau{\cal F}_{+})\cot(\tau{\cal F}_{-})
\nonumber\\
&&\hspace{20pt}
-i\pi\sum_{n=1}^{\infty}\left({\cal {\cal F}_{+}{\cal F}_{-}}
\over n\pi\right)
{\mathrm e}^{-{n\pi\over{\cal F}_{-}}x}
\coth\left({n\pi{\cal F}_{+}\over{\cal F}_{-}}\right)
\label{intEM}
\end{eqnarray}
with ${\cal F}_{+}=F_{+}/\Lambda^{2}$ and 
${\cal F}_{-}=\vert F_{-}\vert/\Lambda^{2}$, obeying 
$0\le{\cal F}_{\pm}<1$. (The sign of $F_{-}$ or 
$F_{\mu\nu}\widetilde F^{\mu\nu}$ is irrelevant because of parity 
invariance.) From (\ref{intEM}), 
${\cal F}_{-}$ is solely responsible for the imaginary part as is 
expected. 
(When ${\cal F}_{+}\rightarrow0$ there remains the imaginary part.) 
Therefore, as was done in the electric field case, we must set the 
condition 
\begin{equation}
{\cal F}_{-}<x
\label{F-x}
\end{equation}
for not having a large imaginary part, which enables us to put 
$\epsilon\rightarrow0$ and again prevent us from talking about 
criticality of the coupling and external fields. 

\begin{figure}[ht]
\centering
(a){\epsfysize=6cm\epsfbox{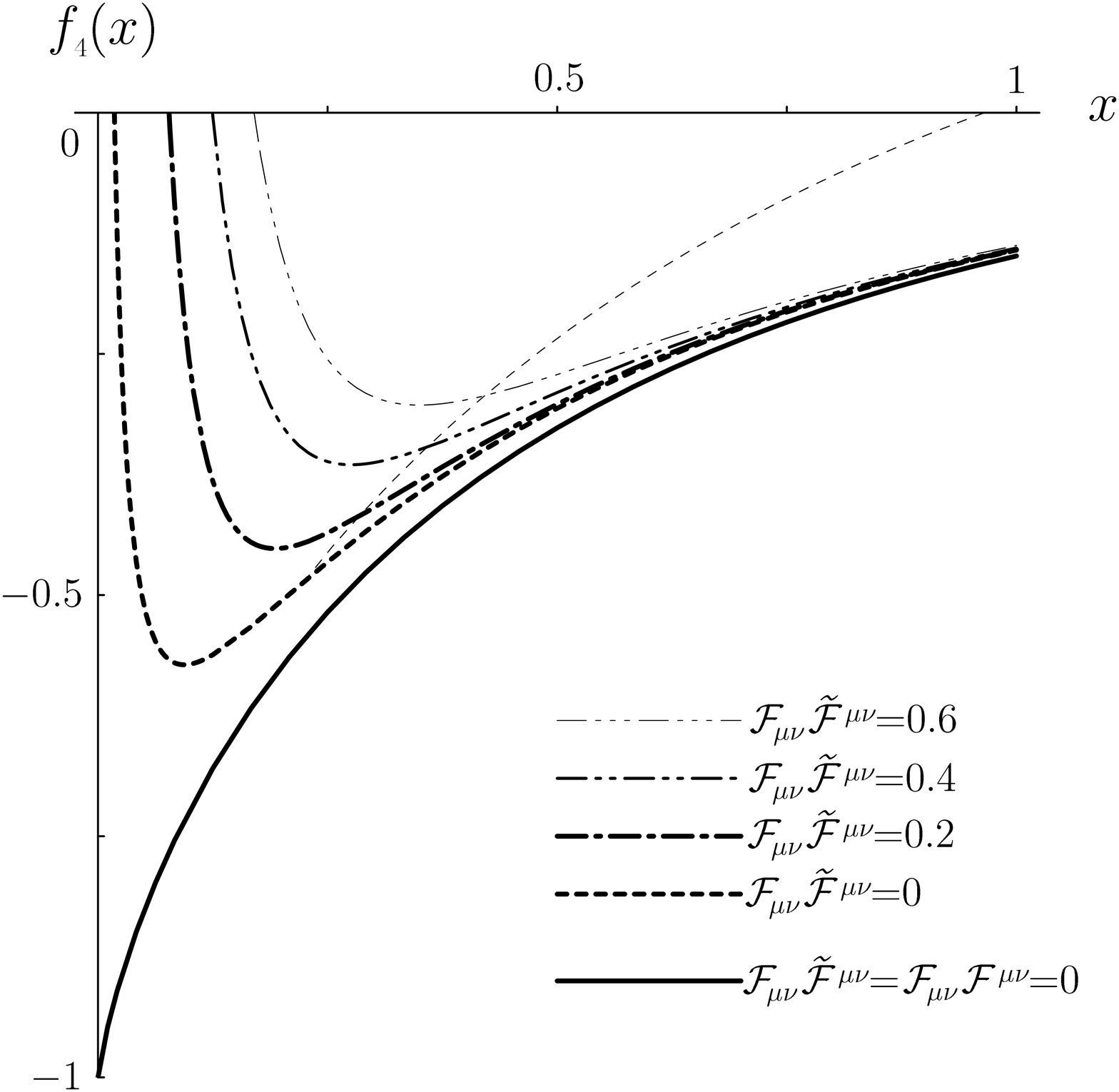}}
(b){\epsfysize=6cm\epsfbox{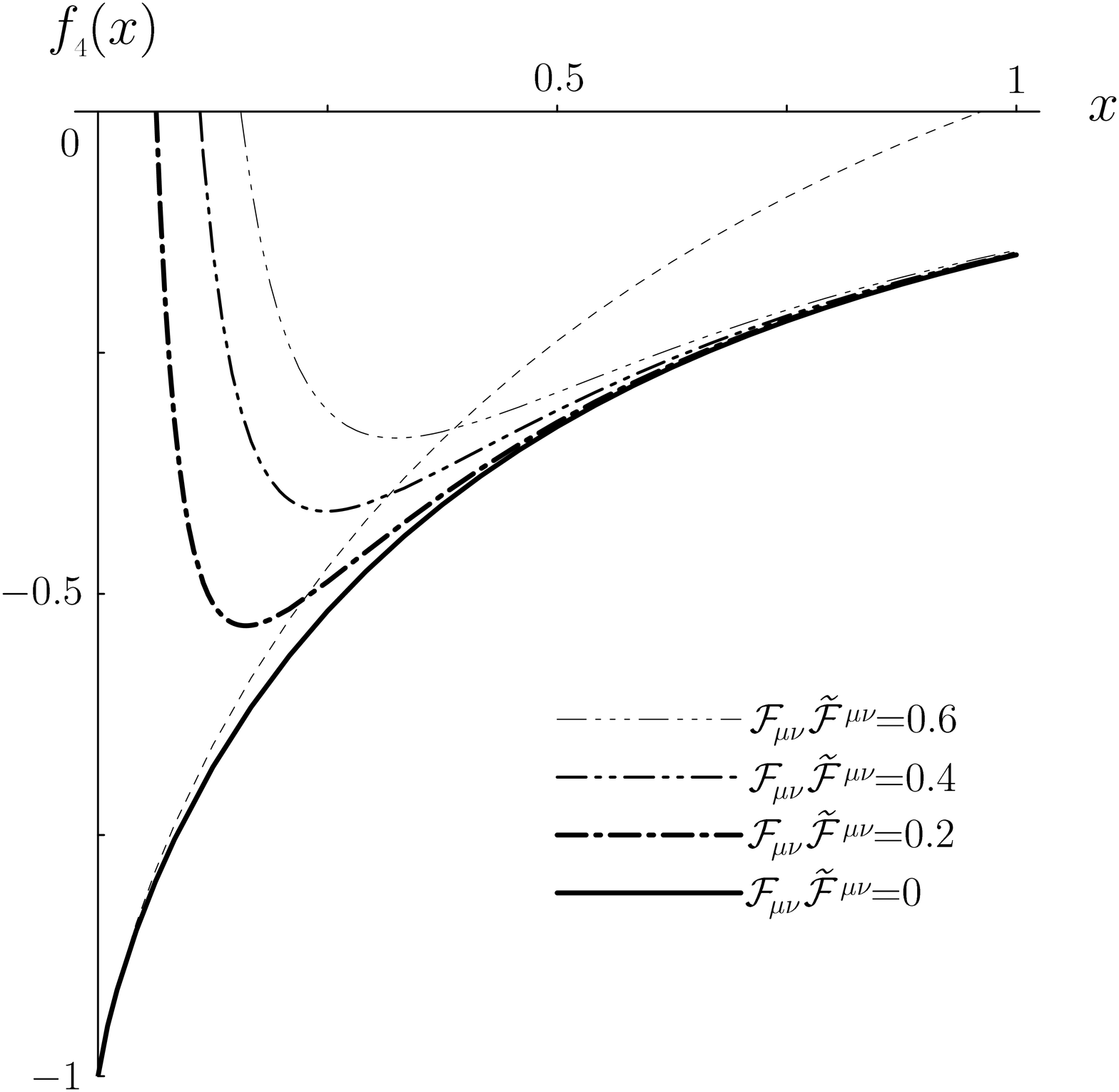}}\\
(c){\epsfysize=6cm\epsfbox{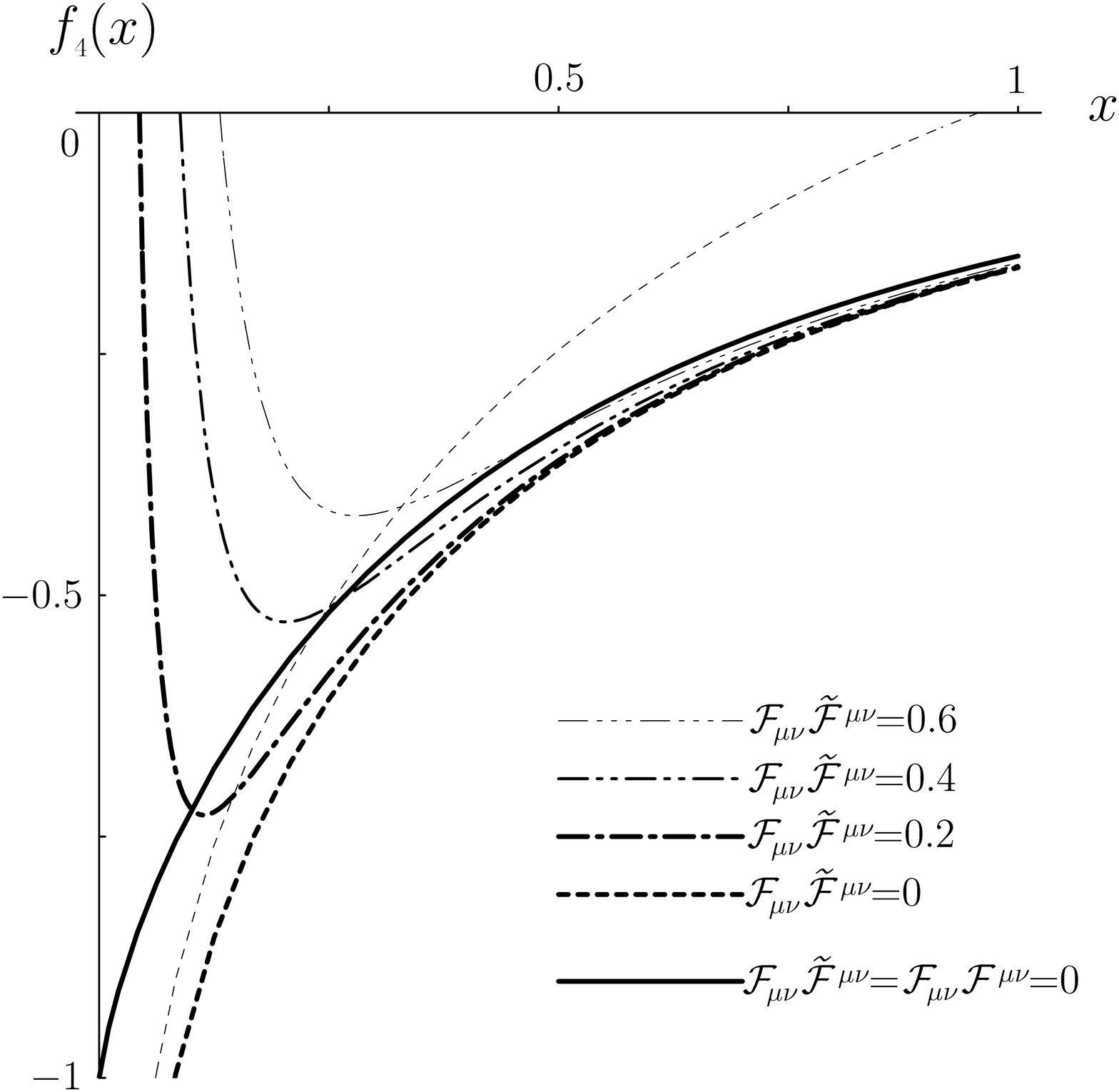}}
\caption{
$f_{4}(x)$ for several fixed 
values of ${\cal F}_{\mu\nu}\widetilde{\cal F}^{\mu\nu}$ 
for the cases, (a) ${\cal F}_{\mu\nu}{\cal F}^{\mu\nu}=0.2$, 
(b) ${\cal F}_{\mu\nu}{\cal F}^{\mu\nu}=0$, and (c) 
${\cal F}_{\mu\nu}{\cal F}^{\mu\nu}=-0.2$. 
The thick line, ${\cal F}_{\mu\nu}\widetilde{\cal F}^{\mu\nu}=
{\cal F}_{\mu\nu}{\cal F}^{\mu\nu}=0$, that is, no external fields, 
is also depicted in comparison. 
The thin dashed line indicates the lower bound of $x$ (\ref{F-x}), 
which is obtained by putting 
${\cal F}_{-}=x$ and 
${\cal F}_{+}=\sqrt{x^{2}+{\cal F}_{\mu\nu}{\cal F}^{\mu\nu}/2}$ in 
$f_{4}(x)$. 
These show that 
${\cal F}_{\mu\nu}\widetilde{\cal F}^{\mu\nu}$ opposes mass generation. 
In (a) ((c)), graphs in the physical region, satisfied with 
(\ref{F-x}), are shifted to right (left) comparing 
to the curves in (b). This reflects the fact that in the magnetic-like 
case, ${\cal F}_{\mu\nu}{\cal F}^{\mu\nu}>0$, the mass goes larger 
while in the electric like case, 
${\cal F}_{\mu\nu}{\cal F}^{\mu\nu}<0$, it goes smaller. }
\label{fxG}
\end{figure}

Now go back to the Euclidean world and regard 
${\cal F}_{+}^{E}$ and ${\cal F}_{-}^{E}$ as 
perturbations as before. Follow the same procedure as in the 
preceding section: first expand the integrand in the left-hand side 
of (\ref{intEM}) such that 
\begin{eqnarray}
\lefteqn{\int_{1}^{\infty}\!\!d\tau\,
\tau^{-3}{\mathrm e}^{-\tau x}
(\tau^2{\cal F}_{+}^{E}{\cal F}_{-}^{E})\coth(\tau{\cal F}_{+}^{E})
\coth(\tau{\cal F}_{-}^{E})}
\nonumber   \\
&\simeq&{\cal F}_{+}^{E}\int_{1}^{\infty}\!\!d\tau\,
\tau^{-2}{\mathrm e}^{-\tau x}
\Big[(\tau{\cal F}_{-}^{E})\coth(\tau{\cal F}_{-}^{E})\Big]
\nonumber\\
&&+\int_{1}^{\infty}\!\!d\tau\,
\tau^{-3}{\mathrm e}^{-\tau (x+2{\cal F}_{+}^{E})}
\left[1+\tau{\cal F}_{+}^{E}+{2\over3}(\tau{\cal F}_{+}^{E})^{2}\right]
\Big[(\tau{\cal F}_{-}^{E})\coth(\tau{\cal F}_{-}^{E})\Big]\ ,
\label{IntMinkowskiG}
\end{eqnarray}
where we have employed the relations (\ref{IntMinkowskiB2}) and 
(\ref{expansion}) developed in the pure magnetic field case, which is 
an extremely good expansion. Then we expand 
$(\tau{\cal F}_{-}^{E})\coth(\tau{\cal F}_{-}^{E})$ by using 
the expansion (\ref{expansionE}), integrate each term, and make 
analytic continuation, ${\cal F}_{-}^{E}\rightarrow-i{\cal F}_{-}$, 
to obtain 
\begin{eqnarray}
\overline{\mbox{\Large$v$}}_{4}(x)&=&{4{\pi}^{2}\over g^{2}\Lambda^{2}}x
+{\cal F}_{+}\left[xEi(-x)
+\left(1-{{\cal F}_{-}^{2}\over3x}\right){\mathrm e}^{-x}\right]
-\left[{x^{2}\over2}+x{\cal F}_{+}
+{{\cal F}_{+}^{2}-{\cal F}_{-}^{2}\over3}\right]Ei(-(x+2{\cal F}_{+}))
\nonumber\\
&&+\left[{1-x\over2}
-{{\cal F}_{+}{\cal F}_{-}^{2}\over3(x+2{\cal F}_{+})}
-{{\cal F}_{+}^{2}{\cal F}_{-}^{2}\over9(x+2{\cal F}_{+})}
\left(1+{1\over x+2{\cal F}_{+}}\right)\right]
{\mathrm e}^{-(x+2{\cal F}_{+})}-{1\over2}\ .
\label{potentialG}
\end{eqnarray}
As was expected, there is no imaginary part. 
The gap equation is then found as
\begin{eqnarray}
\lefteqn{-{4\pi^{2}\over g^{2}\Lambda^{2}}=f_{4}(x)}
\nonumber\\
&=&{\cal F}_{+}\left[Ei(-x)+{{\cal F}_{-}^{2}\over3x}
\left(1+{1\over x}\right){\mathrm e}^{-x}\right]
-(x+{\cal F}_{+})Ei(-(x+2{\cal F}_{+}))
\nonumber\\
&&-\Biggl[1+{{\cal F}_{+}^{2}-{\cal F}_{-}^{2}\over3(x+2{\cal F}_{+})}
-{{\cal F}_{+}{\cal F}_{-}^{2}\over3(x+2{\cal F}_{+})}
\left(1+{1\over x+2{\cal F}_{+}}\right)
\nonumber\\
&&-{{\cal F}_{+}^{2}{\cal F}_{-}^{2}\over9(x+2{\cal F}_{+})}
\left(1+{2\over x+2{\cal F}_{+}}+{2\over(x+2{\cal F}_{+})^{2}}\right)
\Biggr]{\mathrm e}^{-(x+2{\cal F}_{+})}\ .
\end{eqnarray}
The results for the cases, ${\cal F}_{\mu\nu}{\cal F}^{\mu\nu}=0.2$ 
(magnetic-like), ${\cal F}_{\mu\nu}{\cal F}^{\mu\nu}=0$, and 
${\cal F}_{\mu\nu}{\cal F}^{\mu\nu}=-0.2$ (electric-like) 
are shown in Fig.~\ref{fxG} (a)--(c), from which we first notice that 
the dynamical mass becomes always smaller as the magnitude of 
${\cal F}_{\mu\nu}\widetilde{\cal F}^{\mu\nu}$ goes larger. 
This could be 
understood from the following facts: (i) from (\ref{Fpm}), 
${\cal F}_{-}$ is a monotonically increasing function of 
$\vert{\cal F}_{\mu\nu}\widetilde{\cal F}^{\mu\nu}\vert$ 
for a fixed value of 
${\cal F}_{\mu\nu}{\cal F}^{\mu\nu}$. (ii) ${\cal F}_{-}$ is nothing 
but covariant electric field so that ${\cal F}_{-}$ 
becomes larger 
the mass goes smaller according to the discussion on the pure 
electric field case in Sec. \ref{electric}. In this respect it should 
be noted that there is almost no role of ${\cal F}_{+}$, also an 
increasing function of 
$\vert{\cal F}_{\mu\nu}\widetilde{\cal F}^{\mu\nu}\vert$, 
in this phenomena. This might be seen from the form of the imaginary 
part of the effective potential (\ref{intEM}), where ${\cal F}_{-}$ is 
essential. Second, we notice that in Fig.~8 (a) ((c)) graphs in the 
physical region defined by the condition (\ref{F-x}) are shifted 
to the larger (smaller) mass side with 
respect to the curve of no external field, 
${\cal F}_{\mu\nu}{\cal F}^{\mu\nu}
={\cal F}_{\mu\nu}\widetilde{\cal F}^{\mu\nu}=0$, which reflects the 
fact that in the magnetic-like case, 
${\cal F}_{\mu\nu}{\cal F}^{\mu\nu}>0$, the mass goes larger while in 
the electric-like case, ${\cal F}_{\mu\nu}{\cal F}^{\mu\nu}<0$, it 
goes smaller. 
The thin-dashed line indicates the lower bound of $x$ (\ref{F-x}).

\begin{figure}[ht]
\centering\leavevmode
{\epsfysize=6cm\epsfbox{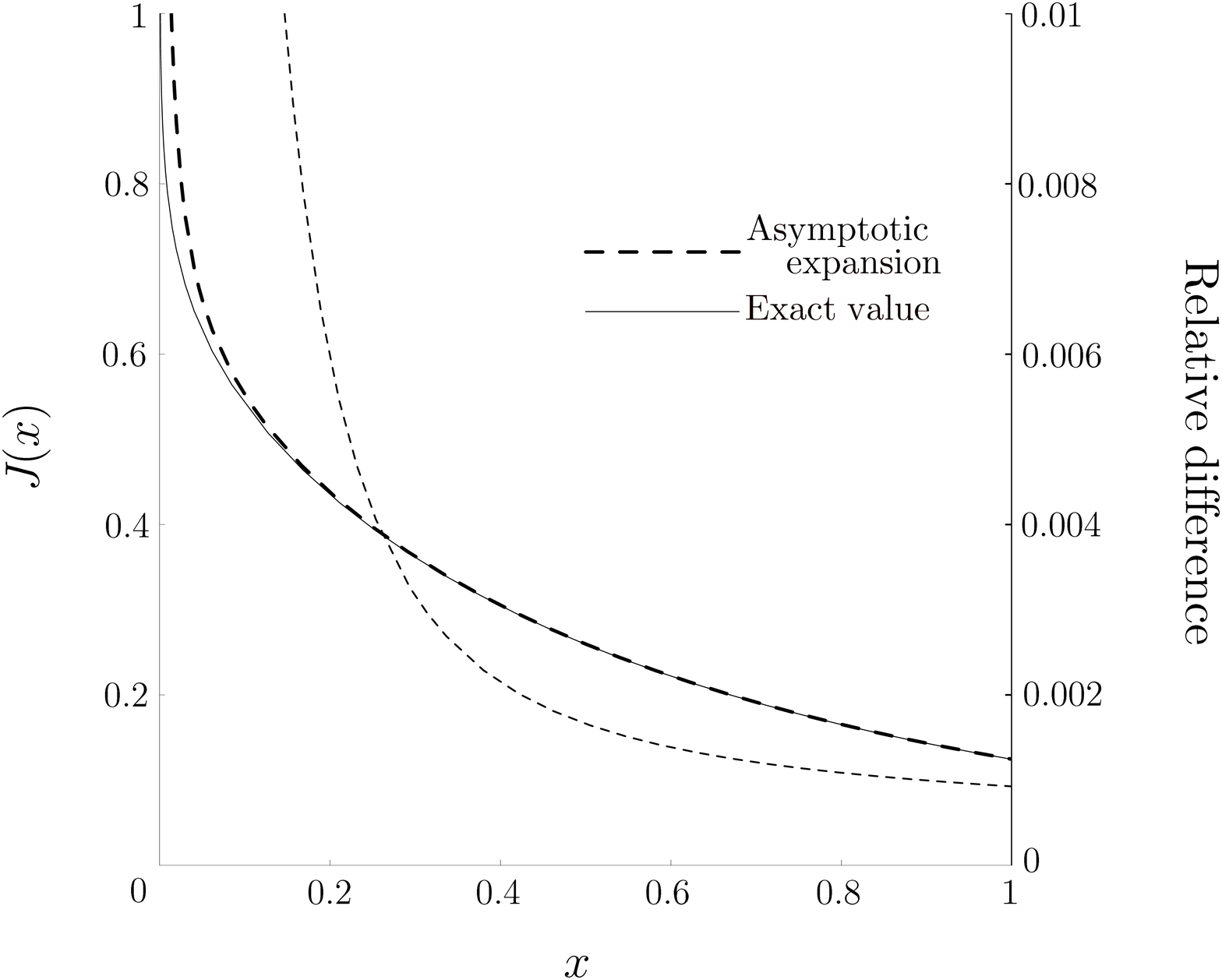}}
\caption{
Plausibility of the asymptotic expansion for general external fields 
in $D=4$: we write the left-hand side of (\ref{intEM}) as 
$J(x)\equiv{\cal F}_{+}{\cal F}_{-}\int_{1}^{\infty}\!\!d\tau\,\tau^{-1}
{\mathrm e}^{-\tau x}\coth(\tau{\cal F}_{+})\coth(\tau{\cal F}_{-})$ 
and put ${\cal F}_{+}=0.4$ and ${\cal F}_{-}=0.2$. The thin line denotes 
the (numerically evaluated) exact value and the dashed line 
represents the asymptotic expansion in the same approximation as the 
one for (\ref{potentialG}). The dotted line denotes the relative 
difference $[(Approx.)-(Exact)]/(Exact)$. Matching is excellent for 
almost all values of $x$ down to ${\cal F}_{-}$. 
}
\label{plauseG}
\end{figure}

It should also be noted that the asymptotic expansion in this case 
matches excellently with the exact value for almost all values of $x$ 
down to ${\cal F}_{-}$. (See Fig.~\ref{plauseG}.)

\section{Conclusion}\label{conclusion}

We examine the NJL model in 2-, 3-, and 4-dimension coupled to 
external constant electromagnetic fields. The results are summarized as 
follows: an electric (magnetic) field reduces (raises) the dynamically 
generated masses, that is, electric fields oppose
$\chi$SB, while magnetic fields enhance it. 

In the case for a pure magnetic field (in 3- and 4-dimension), 
we obtain a well-defined 
effective potential with a UV cut-off $\Lambda^{2}$ as well as an IR 
cut-off $\epsilon$. When $\epsilon$ is nonzero, we obtain a non-zero 
critical coupling which, however, goes to zero as 
$\epsilon\rightarrow0$. We thus reconfirm that {\it the critical coupling 
is zero for any infinitesimal magnetic field}. 

We find that the effective potential 
has the imaginary part in the electric case defined by condition 
$F_{\mu\nu}F^{\mu\nu}<0$ in 
2- and 3-dimension and $F_{\mu\nu}F^{\mu\nu}<0$ 
or $F_{\mu\nu}\widetilde F^{\mu\nu}\ne0$ in 4-dimension. 
In these cases, imaginary parts should be small and negligible, so 
that the condition, ${\cal E}<x$ as well as ${\cal F}_{-}<x$ 
in 4-dimension, has been obtained, which enables us to put the IR 
cut-off $\epsilon$ zero. Moreover, this condition prevents us 
from selecting out the critical quantities, such as critical 
couplings or critical fields, since those are defined  through the 
transition from a massive to a massless state or vice versa. 

However the imaginary part cannot be seen under the perturbation 
theory with respect to the external fields. This perturbative 
expansion, realized as an asymptotic expansion, excellently matches 
with the exact value by adopting only the first few terms. 
We have assumed that the external fields are 
constant, so that we can calculate the functional determinant exactly. 
If, however, the fields depend on space-time, we cannot calculate it 
without approximation. In a usual approach, a weak field approximation 
is employed, which would therefore be a fairly good 
expansion according to our analysis. 

In our present work, we can ignore the chiral anomaly, which is 
trivial in the abelian case but 
indispensable to investigate the reality of the dynamics 
of QCD in the low energy region. It is also necessary to work 
with non-constant gauge field, such as the instanton configuration. 
Generalization of the present 
work to a non-abelian gauge field and including the effect of the anomaly 
is our next step.

\appendix
\section{Calculation of $G_{D}(\tau F)$}\label{appendixa}

In this appendix the calculation of $G_{D}(F)$ with a real 
antisymmetric tensor $F_{\mu\nu}$ is explicitly 
given. 

The definition for $G_{D}(F)$ is 
\begin{equation}
G_{D}(F)\equiv
{{\mathrm{tr}}({\mathrm e}^{{1\over2}\sigma_{\mu\nu}{F}_{\mu\nu}})
\over{\mathrm{tr}}{\bf1}}
\left[\det\left({\sin{F}\over{F}}\right)\right]^{-{1\over2}}\ ,
\label{Gdef}
\end{equation}
where ${\mathrm tr}$ is for gamma matrices and ${\det}$ is for 
$F_{\mu\nu}$ as a $D\times D$ matrix. The result is 
\begin{equation}
G_{D}(F)=\left\{
\begin{array}{lcl}
{F}_{D}\coth{F}_{D}&{\mathrm{for}}&D=2,3\ ,\\
\noalign{\vspace{1ex}}
{F}_{+}{F}_{-}\coth{F}_{+}\coth{F}_{-}
&{\mathrm{for}}&D=4\ ,
\end{array}\right.
\label{Gresult}
\end{equation}
with  
\begin{equation}
\left\{
\begin{array}{lcl}
{F}_{2}&=&E\ ,\quad{F}_{3}=\sqrt{B^{2}+{\bbox E}^{2}}\ ,\\
{F}_{\pm}&=&\left\{\vert{\bbox{B+E}}\vert\pm
\vert{\bbox{B-E}}\vert\right\}/2\ ,
\end{array}\right.\label{Fds}
\end{equation}
where $E$ (or ${\bbox E}$) and $B$ (or ${\bbox B}$) denote an 
electric field and a magnetic field respectively. 

The derivations of (\ref{Gresult}) and (\ref{Fds}) are shown as follows: 
in $D=2$ case, they can easily be obtained with 
\begin{equation}
\gamma_{\mu}=\sigma_{\mu}\,,\quad\gamma_{5}
=-i\gamma_{1}\gamma_{2}\,,\quad
\sigma_{\mu\nu}\equiv[\gamma_{\mu},\gamma_{\nu}]
/2i=\epsilon_{\mu\nu}\sigma_{3}\,;\quad\mu,\nu=1,2
\end{equation}
and with
\begin{equation}
(F_{\mu\nu})=\left(\begin{array}{cc}0&E\\-E&0\end{array}\right)\ ,
\end{equation}
which can be diagonalized into $iE{\mathrm diag}[1,-1]$. 
Therefore 
\begin{equation}
G_{2}(F)=
{{\mathrm{tr}}({\mathrm e}^{\sigma_{3}E})
\over{\mathrm{tr}}{\bf1}}
\left[\det\left({\sinh E\over E}\right)\right]^{-{1\over2}}
=E\coth E\ .
\end{equation}
In $D=3$ case, 
since $F_{\mu\nu}$ in the $3$-dimensional space-time is expressed as 
\begin{equation}
(F_{\mu\nu})=\left(
\begin{array}{ccc}0&B&E_{1}\\-B&0&E_{2}\\-E_{1}&-E_{2}&0\end{array}
\right)=i(E_{2}T_{1}-E_{1}T_{2}+BT_{3})\ ,
\end{equation}
where $T_{i}(i=1,2,3)$ is a basis of the $SO(3)$ algebra in the 
adjoint representation, 
it can be diagonalized into 
$i\sqrt{B^{2}+{\bbox E}^{2}}{\mathrm{diag}}[1,0,-1]$. Therefore 
\begin{equation}
\left[\det\left({\sin F\over F}\right)\right]^{-{1\over2}}
={{F}_{3}\over\sinh {F}_{3}}\ .
\label{det3}
\end{equation}

The calculation of the trace part of (\ref{Gresult}) can be done with 
the use of gamma matrices in 3-dimension, given by 
\begin{equation}
\gamma_{\mu}=\left(\begin{array}{cc}\sigma_{\mu}&0\\
0&-\sigma_{\mu}\end{array}\right)\,,\quad\sigma_{\mu\nu}={1\over2i}
[\gamma_{\mu},\gamma_{\nu}]=\epsilon_{\mu\nu\rho}\sigma_{\rho}\,;\quad
\mu,\nu,\rho=1,2,3\,.
\end{equation}
Therefore 
\begin{equation}
{1\over2}\sigma_{\mu\nu}F_{\mu\nu}={1\over2}
F_{\mu\nu}\epsilon_{\mu\nu\rho}\left(
\begin{array}{cc}\sigma_{\rho}&0\\0&\sigma_{\rho}\end{array}\right)
=\left(\begin{array}{cc}E_{2}\sigma_{1}-E_{1}\sigma_{2}+B\sigma_{3}&0\\
0&E_{2}\sigma_{1}-E_{1}\sigma_{2}+B\sigma_{3}\end{array}\right)\ ,
\end{equation}
which can be diagonalized as 
$\sqrt{B^{2}+{\bbox{E}}^{2}}{\mathrm{diag}}[1,-1,1,-1]$, yielding to 
\begin{equation}
{{\mathrm{tr}}({\mathrm e}^{{1\over2}\sigma_{\mu\nu}{\cal F}_{\mu\nu}})\over
{\mathrm{tr}}{\bf1}}=\cosh{F}_{3}\ .
\label{tr3}
\end{equation}
Combining (\ref{det3}) and (\ref{tr3}), we obtain (\ref{Gresult}) and 
(\ref{Fds}) in $D=3$ case. 

In $D=4$ case, $F_{\mu\nu}$ is expressed as 
\begin{equation}
F_{\mu\nu}=\left(\begin{array}{cccc}0&B_{3}&-B_{2}&E_{1}\\
-B_{3}&0&B_{1}&E_{2}\\B_{2}&-B_{1}&0&E_{3}\\-E_{1}&-E_{2}&-E_{3}&0
\end{array}\right)_{\mu\nu}
=i\left({\bbox{B\cdot M}}+{\bbox{E\cdot N}}\right)_{\mu\nu}\ ,
\end{equation}
where $M_{i}$ and $N_{i}$ (i=1,2,3) is a basis of the 
$SO(4)$ algebra 
in the fundamental representation. Since $M_{i}$ and $N_{i}$ are 
decomposed into the subalgebra
\begin{equation}
{\bbox J}\equiv{1\over2}({\bbox{M+N}})\,,\quad
{\bbox K}\equiv{1\over2}({\bbox{M-N}})\ ,
\end{equation}
satisfying 
\begin{equation}
[J_{i},J_{j}]=i\epsilon_{ijk}J_{k}\,,\quad
[K_{i},K_{j}]=i\epsilon_{ijk}K_{k}\,,\quad
[J_{i},K_{j}]=0\ ,
\label{algebra}
\end{equation}
$F_{\mu\nu}$, now expressed as 
$i\left\{{\bbox{(B+E)\cdot J+(B-E)\cdot K}}\right\}_{\mu\nu}$, 
can be diagonalized into $i{\mathrm{diag}}[F_{+},-F_{+},F_{-},-F_{-}]$. 
Therefore 
\begin{equation}
\left[\det\left({\sin F\over F}\right)
\right]^{-{1\over2}}
={{F}_{+}{F}_{-}
\over\sinh{F}_{+}\sinh{F}_{-}}\ .
\label{det4}
\end{equation}
As for the trace, in order to diagonalize 
$\sigma_{\mu\nu}F_{\mu\nu}$, we first define 
\begin{equation}
{\cal M}_{\mu}\equiv{1\over4}\epsilon_{\mu\nu\rho}\sigma_{\nu\rho}\,,\quad
{\cal N}_{\mu}\equiv{1\over2}\sigma_{\mu4}\,;\quad\mu,\nu,\rho=1,2,3\ .
\end{equation}
Similar to ${\bbox M}$ and ${\bbox N}$, ${\cal M}_{i}$ and ${\cal N}_{i}$ can 
be decomposed into the subalgebra, 
\begin{equation}
{\cal J}_{i}\equiv{1\over2}({\cal M}_{i}+{\cal N}_{i})\,,\quad
{\cal K}_{i}\equiv{1\over2}({\cal M}_{i}-{\cal N}_{i})\ ,
\end{equation}
which also satisfies the same algebra as (\ref{algebra}). Thus
\begin{equation}
{1\over2}\sigma_{\mu\nu}F_{\mu\nu}=2(B_{i}{\cal M}_{i}+E_{i}{\cal N}_{i})
=2\left(({\bbox{B+E}})_{i}{\cal J}_{i}+({\bbox{B-E}})_{i}{\cal K}_{i}
\right)
\end{equation}
which can be diagonalized into ${\mathrm{diag}}[\vert{\bbox{B+E}}\vert,
-\vert{\bbox{B+E}}\vert,\vert{\bbox{B-E}}\vert,-\vert{\bbox{B+E}}\vert]$. 
Hence 
\begin{equation}
{{\mathrm{tr}}({\mathrm e}^{{1\over2}\sigma_{\mu\nu}{F}_{\mu\nu}})\over
{\mathrm{tr}}{\bf1}}={1\over2}\cosh\vert{\bbox{B+E}}\vert
+{1\over2}\cosh\vert{\bbox{B-E}}\vert
=\cosh{F}_{+}\cosh{F}_{-}\ .
\label{tr4}
\end{equation}
Combining (\ref{det4}) and (\ref{tr4}), we finally obtain 
(\ref{Gresult}) and (\ref{Fds}) for $D=4$ case.  

\vspace{5ex}
\centerline{\large\bf Acknowledgment}
\vspace{3ex}
\noindent The authors thank to K.~Inoue, K.~Harada, and 
T.~Kugo for valuable discussions. 

%

\end{document}